\documentclass[a4paper,usenatbib,useAMS]{mnras}

\usepackage{graphicx}	% Including figure files
\usepackage{amsmath}	% Advanced maths commands
\usepackage{amssymb}	% Extra maths symbols
\usepackage{multicol}        % Multi-column entries in tables
\usepackage{bm}		% Bold maths symbols, including upright Greek
\usepackage{pdflscape}	% Landscape pages
\usepackage{indentfirst}
\usepackage{hyperref}

\usepackage[T1]{fontenc}
\usepackage{ae,aecompl}

\usepackage{newtxtext,newtxmath}
\title[Intrinsic alignments evolution]{The evolution of galaxy intrinsic alignments in the \texttt{MassiveBlack II} universe}

\author[Bhomick et al.]{
Aklant K. Bhowmick$^{1}$,
Yingzhang Chen$^{1}$, Ananth Tenneti$^{1}$, Tiziana Di Matteo$^{1}$,
\newauthor
Rachel Mandelbaum$^{1}$
\\
$^{1}$McWilliams Center for Cosmology, Department of Physics, Carnegie Mellon University, Pittsburgh, PA 15213, USA}

\pubyear{2018}

\begin{document}
\label{firstpage}
\pagerange{\pageref{firstpage}--\pageref{lastpage}}
\maketitle

\begin{abstract}
We investigate the redshift evolution of the intrinsic alignments (IA) of galaxies in the \texttt{MassiveBlackII} (MBII) simulation. We select galaxy samples above fixed subhalo mass cuts~($M_h>10^{11,12,13}~M_{\odot}/h$) at $z=0.6$ and trace their progenitors to $z=3$ along their merger trees. Dark matter components of $z=0.6$ galaxies are more spherical than their progenitors while stellar matter components tend to be less spherical than their progenitors. The distribution of the galaxy-subhalo misalignment angle peaks at $\sim10~\mathrm{deg}$ with a mild increase with time. The evolution of the ellipticity-direction~(ED) correlation amplitude $\omega(r)$ of galaxies (which quantifies the tendency of galaxies to preferentially point towards surrounding matter overdensities) is governed by the evolution in the alignment of underlying dark matter~(DM) subhaloes to the matter density of field, as well as the alignment between galaxies and their DM subhaloes. At scales  $\sim1~\mathrm{cMpc}/h$, 
 the alignment between DM subhaloes and matter overdensity gets suppressed with time, whereas the alignment between galaxies and DM subhaloes is enhanced. These competing tendencies lead to a complex redshift evolution of $\omega(r)$ for galaxies at $\sim1~\mathrm{cMpc}/h$. At scales $>1~\mathrm{cMpc}/h$, alignment between DM subhaloes and matter overdensity does not evolve significantly; the evolution of the galaxy-subhalo misalignment therefore leads to an increase in  $\omega(r)$ for galaxies by a factor of $\sim4$ from $z=3$ to $0.6$ at scales $>1~\mathrm{cMpc}/h$. The balance between competing physical effects is scale dependant, leading to different conclusions at much smaller scales~($\sim0.1~\mathrm{Mpc}/h$).  %\rachel{I think it's too much to go into this much detail about both larger and smaller scales.  I would recommend simplifying everything from `At $\sim 0.1$\dots misalignment' to something like `'} 

\end{abstract}

%We also compare our results to sample selection applied in previous works where galaxy samples~(\texttt{SAMPLE-MCUT}) were based on fixed subhalo mass cut~($M_h>10^{11,12,13}~M_{\odot}/h$) applied at all redshifts~(thereby including the effect of sample selection in the redshift evolution). Notably, the sphericities and their redshift evolution of \texttt{SAMPLE-TREE} and \texttt{SAMPLE-MCUT} galaxies do not significantly differ~($\lesssim 10\%$) even at the highest redshift~($z=3$) where \texttt{SAMPLE-TREE} galaxies are significantly less massive than \texttt{SAMPLE-MCUT}. This is because at $z\gtrsim1.5$, progenitors of $z\sim0.6$: $M_h\gtrsim10^{11}~M_{\odot}/h$ galaxies are significantly less spherical~(on an average) than a \textit{randomly selected} galaxy of similar subhalo mass and redshift. The ED correlation $\omega(r)$ for \texttt{SAMPLE-TREE} galaxies is suppressed~(due to decreasing subhalo masses) compared to \texttt{SAMPLE-MCUT}; the suppression increases with redshift (to factors $\sim3-4$ at $z=3$).

%This is an extension of the previous paper on IA of MBII galaxies \citep{tenneti2015intrinsic} which considers objects selected by fixed subhalo mass ($M_H>10^{11,12,13}~M_{\odot}/h$) at all redshifts.\aklant{I am editing this. Previous version is commented out in the .tex file} 

\begin{keywords}
%gravitational lensing, intrinsic alignment, misalignment angle, correlation function \ananth{change keywords : 
methods: numerical -- hydrodynamics -- gravitational lensing: weak -- galaxies: star formation
% Rachel: I fixed this to use the ones from the list that Ananth suggested, instead of the original ones (which were not on the allowed list).
\end{keywords}

%%%%%%%%%%%%%%%%%%%%%%%%%%%%%%%%%%%%%%%%%%%%%%%%%%

%%%%%%%%%%%%%%%%% BODY OF PAPER %%%%%%%%%%%%%%%%%%

% The MNRAS class isn't designed to include a table of contents, but for this document one is useful.
% I therefore have to do some kludging to make it work without masses of blank space.

%\rachel{Comments on abstract: this should be high-level and stand on its own, so don't introduce jargon like SAMPLE-TREE.  You should leave the definition out of the abstract, and make sure to introduce it in the text.}

\section{Introduction}
The shapes and orientations of galaxies have an intrinsic correlation with respect to those of nearby galaxies and the overall matter distribution; this effect is known as galaxy intrinsic alignments \citep[IA; see][and references therein for review]{troxel2015intrinsic,joachimi2015galaxy,kiessling2015galaxy,kirk2015galaxy}. The importance of IA is two fold: 1) IA emerges as a natural outcome of the current paradigm of galaxy formation in the $\Lambda$CDM cosmological model, as emphasized also in state-of-the-art cosmological hydrodynamic simulations that include direct modeling of galaxy formation ~\citep[e.g.,][]{tenneti2014galaxy,2015MNRAS.454.3328V,2015MNRAS.454.2736C,2017MNRAS.468..790H}. %,tenneti2015intrinsic
IA is therefore a promising probe for galaxy formation physics. 2) If not properly modeled and removed, IA is a significant source of systematic bias in inferring cosmological parameters in weak lensing studies~\citep{2016MNRAS.456..207K}. Many of the upcoming surveys like the Large Synoptic Survey Telescope \citep[LSST;][]{2008arXiv0805.2366I,abell2009lsst}, Euclid \citep{laureijs2011euclid}, and the Wide-Field Infrared Survey Telescope (WFIRST; \citealt{spergel2015wide}) aim to determine the dark energy equation of state to very high precision using weak lensing, and IA is one of the major sources of astrophysical systematic uncertainty for such studies \citep{2018ARA&A..56..393M}.  The existence of IA in galaxies with correlations  out to 100~$h^{-1}$Mpc scales has been firmly established in observational data \citep[e.g.,][]{2006MNRAS.367..611M, 2007MNRAS.381.1197H,2011A&A...527A..26J,singh2015intrinsic}. %singh2016intrinsic,huang2017intrinsic}
 An understanding of intrinsic alignments and their scaling with galaxy mass and redshift is therefore crucial to mitigating this effect in weak lensing studies, and is also 
a good diagnostic for galaxy formation physics.

Intrinsic alignments have been studied using analytical methods such as the linear~\citep{2001MNRAS.320L...7C}, the nonlinear alignment model \citep{2007NJPh....9..444B}, and the full tidal alignment model \citep{blazek2015tidal}. While these methods are easy to implement while also requiring few computational resources, they inevitably rely on assumptions about the alignment of galaxies and the underlying tidal field. This limitation can be overcome by state-of-the-art cosmological hydrodynamic simulations~\citep[e.g.,][]{2014MNRAS.444.1453D,2015MNRAS.446..521S,khandai2015massiveblack,vogelsberger2014introducing}, which can directly probe the impact of galaxy formation physics on the shapes and alignments of galaxies and the relation to their dark matter counterparts~(halos/subhalos)
and the tidal fields themselves. Therefore, in recent years galaxy shapes and alignments have been extensively studied using hydrodynamic simulations \citep[e.g.,][]{2015MNRAS.454.2736C,tenneti2016intrinsic,2017MNRAS.472.1163C,2017MNRAS.468..790H}   

An important step towards understanding galaxy intrinsic alignments is to study their redshift evolution. This has been initiated by a series of works~\citep{tenneti2015intrinsic} %tenneti2016intrinsic} 
using the \texttt{MassiveBlackII} (MBII) hydrodynamic simulation \citep{khandai2015massiveblack}, including a detailed study of the redshift evolution of galaxy shapes, alignment with respect to host halo/subhalo, and associated shape-density correlation functions.
A noteworthy feature of these works was that the sampling of galaxies was based on fixed subhalo mass cut~($\gtrsim 10^{11},~10^{12},~10^{13}~M_{\odot}/h$) at each redshift~(from $z\sim0.06-1$); this is somewhat representative of cuts in observed galaxy samples in properties such as stellar mass or magnitude, which are known to correlate with the host subhalo mass. However, with such an approach, the resulting redshift evolution may be dominated by the effects of sample selection. In order to study the \textit{intrinsic} redshift evolution~(i.e. separated from the effects of sample selection), we must select samples of galaxies at a given redshift and trace their progenitors to higher redshifts.   

In this work, we study the redshift evolution of IA properties of MBII galaxies by making subhalo mass cuts at a single fixed redshift~($z\sim0.6$) and then tracing the properties of their progenitors along a merger tree. In Section~\ref{S:methods}, we outline the basic methodology and definitions. In Section ~\ref{S:results}, we study the redshift evolution of galaxy properties~(axis ratios, galaxy-subhalo misalignment angle and density-shape correlation functions) on the merger tree. We summarize our key results in Section~\ref{S:conclusions}.

\section{Methods} \label{S:methods}
\subsection{MassiveBlack-II simulation}
\begin{figure}
 \includegraphics[width=80mm]{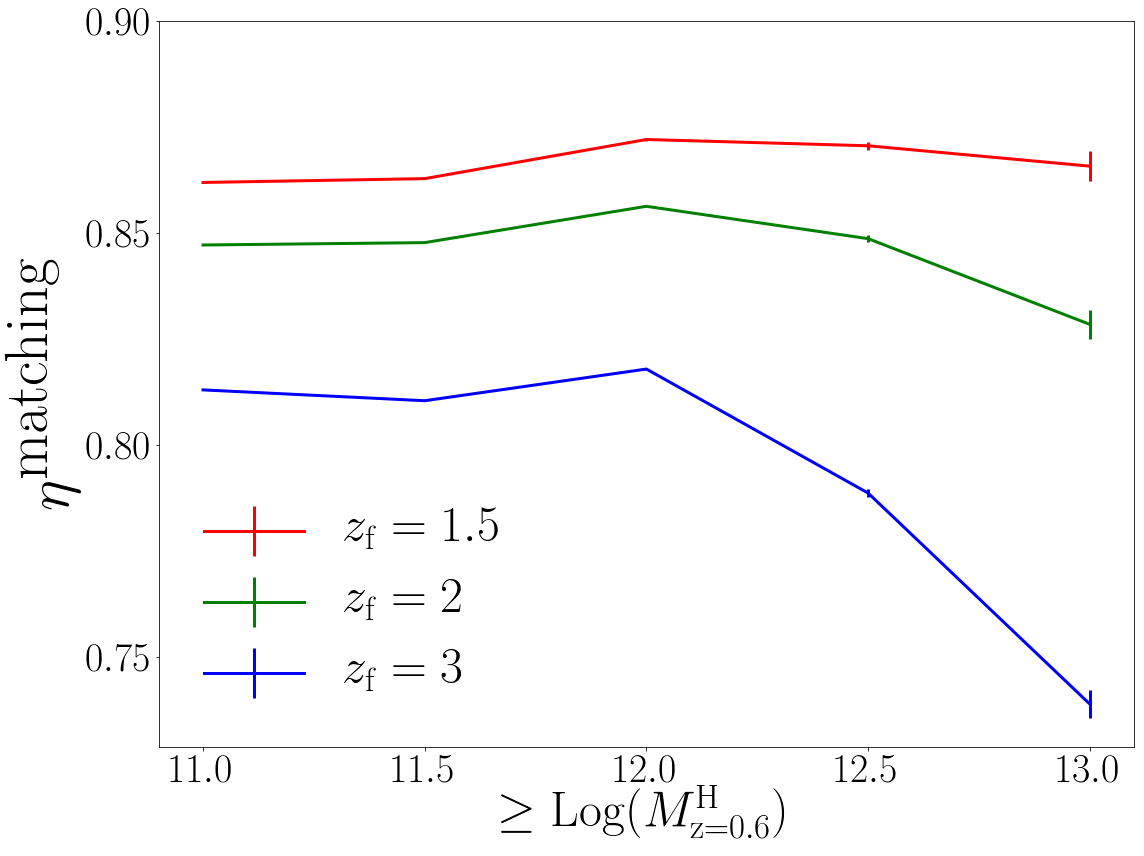}
 \caption{$\eta_{\mathrm{matching}}$ is the \textit{matching efficiency}, i.e.\ the ratio between the number of \texttt{SUBFIND} trees with respect to the original number of \texttt{ROCKSTAR} trees~(before matching \texttt{ROCKSTAR} and \texttt{SUBFIND} trees). $1-\eta_{\mathrm{matching}}$ therefore is the fraction of \texttt{ROCKSTAR} trees lost because we could not find a corresponding \texttt{SUBFIND} tree to match with.  ``$\geq\log(M^H_{z=0.6})$'' is the threshold subhalo mass of galaxies selected at $z=0.6$; $z_f$ is the maximum redshift up to which their progenitors are traced~(starting from $z_i=0.6$). }
 \label{matching_efficiency}
\end{figure}

\begin{figure}
 \includegraphics[width=7.5cm]{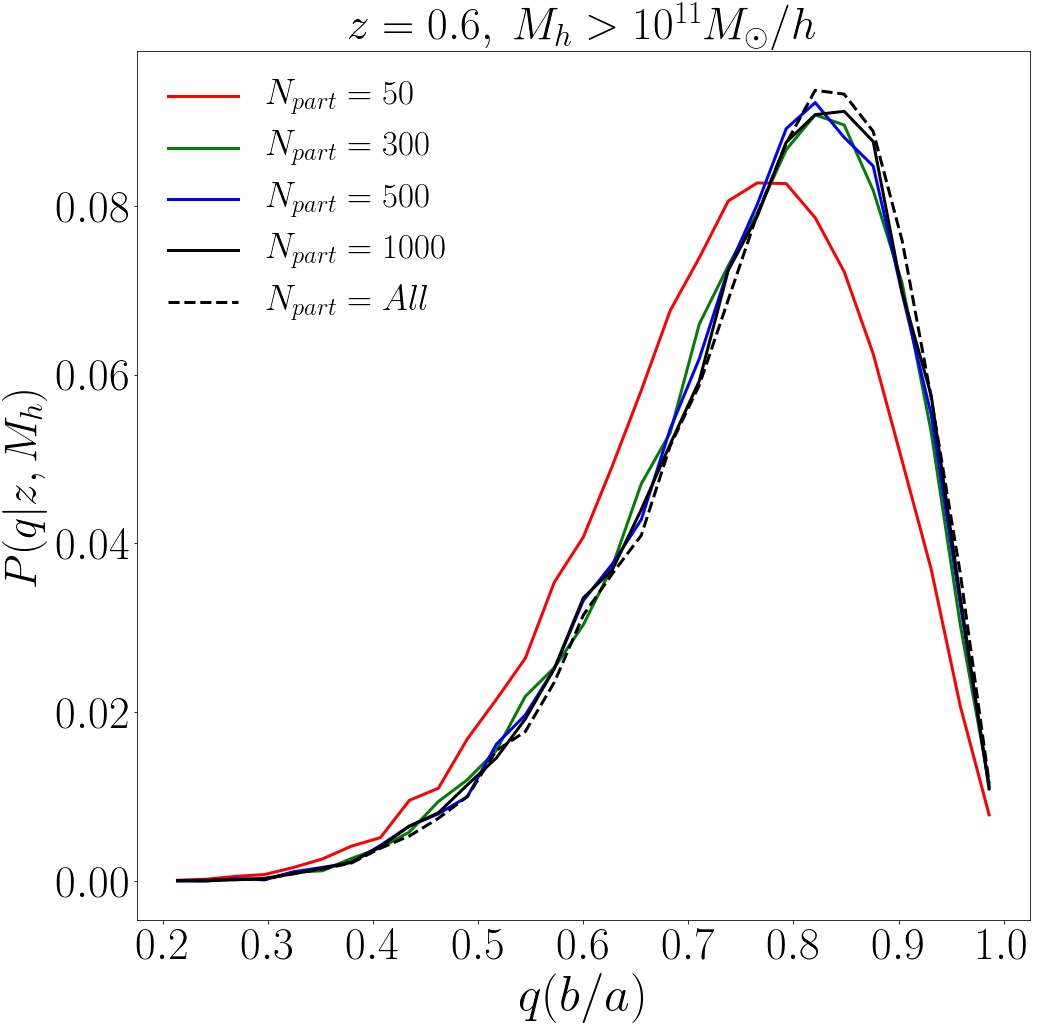}
 \caption{\textbf{Shape convergence test:} Normalized histograms of $q=\frac{b}{a}$ of the dark matter component of \texttt{SAMPLE-TREE} galaxies at $z = 0.6$. We show the comparison between shapes determined using all particles in the subhalo with those obtained using a random subsample of $N_{\mathrm{part}}=50, 100, 300, 1000$ particles in the subhalo.}
 \label{shape_convergence}
\end{figure}

%\begin{figure}
% \includegraphics[width=80mm]{methods.png}
% \caption{Histogram of the axis ratio $q=\frac{b}{a}$ of dark matter subhalo with mass $>10^{11}M_{\odot}h^{-1}$ on the merger tree at $z=0.6$ distribution under different shape definitions.}
%\label{fig:m_angle}
%\end{figure}

\begin{figure*}
 \includegraphics[width=\textwidth]{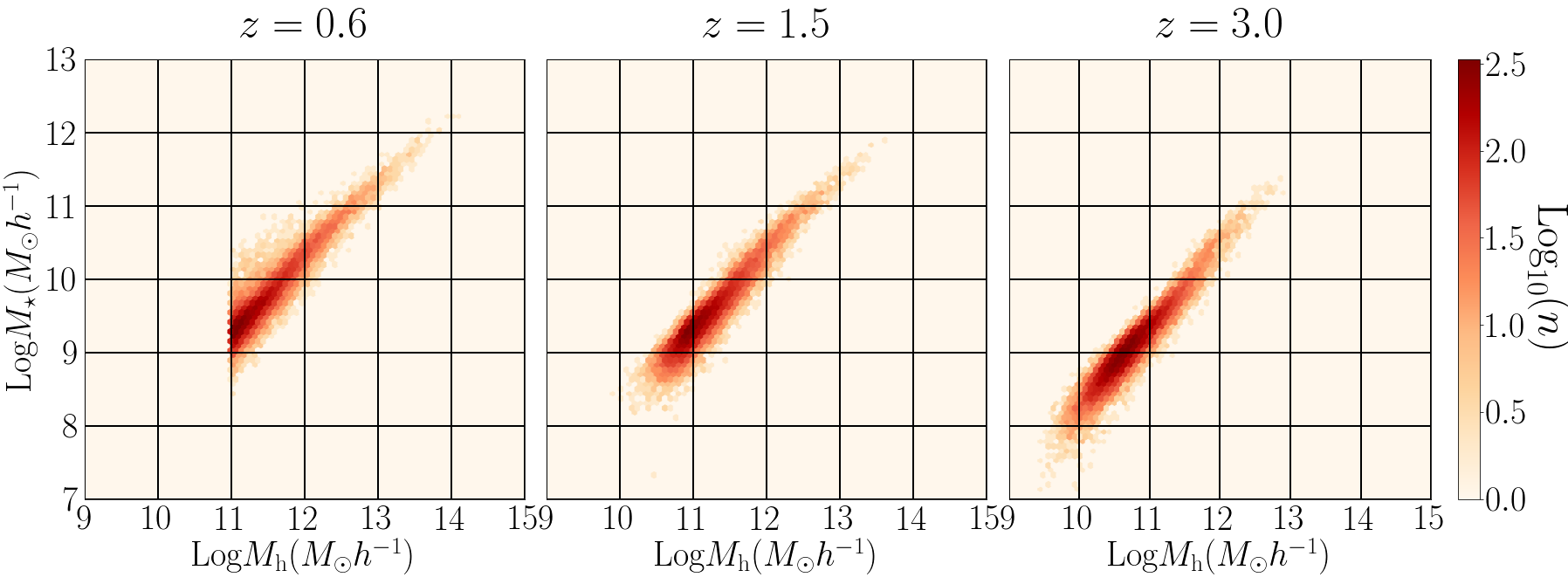}
 \caption{The 2D histograms show the dark matter mass~($M_h$) versus stellar mass~($M_*$) relation of galaxies~(and dark matter subhaloes) on 27942 trees corresponding to $M_{h}>10^{11}M_{\odot}/h$ galaxies at $z=0.6$~(leftmost panel) and their main progenitors at $z=1.5$~(middle panel) and $z=3$~(rightmost panel). %\rachel{Say what the color scale is: log of number of galaxies?}
 }
 \label{SM_HM_fig}
\end{figure*}
We briefly describe \texttt{MassiveBlack-II} (MB-II), which is a state-of-the-art cosmological hydrodynamic simulation of structure formation \citep{khandai2015massiveblack}. MB-II is evolved from $z=159$ to $z=0.06$ in a cubic periodic box of comoving volume $V_{\mathrm{box}}=(100~h^{-1}\mathrm{Mpc})^3$ and a gravitational smoothing length of $\epsilon = 1.85 h^{-1}~\mathrm{kpc}$. The box contains $2\times 1792^3$ particles (dark matter+gas). The mass of a single dark matter particle and a single gas particle is $m_{\mathrm{DM}}=1.1\times 10^7 h^{-1} M_{\odot} $ and $m_{\mathrm{gas}}=2.2\times 10^6 h^{-1} M_{\odot}$ respectively. The cosmological parameters used in the simulation are based on WMAP7 \citep{komatsu2011astrophys} with amplitude of matter fluctuations $\sigma_8 = 0.816$, spectral index $n_s = 0.96$, mass density parameter $\Omega_m = 0.275$, cosmological constant density parameter $\Omega_{\Lambda} = 0.725$, baryon density parameter $\Omega_{b} = 0.046$, and Hubble parameter $h = 0.702$. Halos are identified using a friends-of-friends (FOF) halo finder \citep{davis1985evolution} with a linking length of 0.2 times the mean particle separation.
% "linking length" is not software, so it doesn't need special fonts

\subsection{Galaxy identification}
Here we describe how galaxies are identified in MBII. Galaxies are defined to be the stellar component of \textit{subhalos}, which are locally overdense, self-bound particle groups within a larger parent group~(FOF halo)
The subhalo catalogs are generated using the substructure finder \texttt{SUBFIND} on the halo catalogs. In \texttt{SUBFIND}, for each particle in the parent group, a local density is estimated using the positions of a prescribed number of nearest neighbours. After identifying the local peaks in density field, it rebuilds the parent group by adding particles in the order of decreasing density. In doing so, a saddle point is eventually reached which connects two disjoint overdense regions. The smaller structure is then identified as a candidate substructure. For further implementation details, see the original paper \citep{springel2001populating}.

%\subsection{Determining galaxy shapes}

%Based on convergence tests in Appendix A, we only analyze the measured stellar shapes if there are more than $1000$ dark matter and stellar matter particles. The subhalo mass will decrease at higher redshifts, so we impose a mass cut of $M_{halo}=10^{11}M_{\odot}/h$ at $z=0.6$ to assure that most of our studied samples satisfy the convergence condition.

\subsection{Constructing the galaxy merger tree}
\label{merger_tree_sec}
%This merger history can also be in cosmological simulations and represented in the form of \textit{merger trees}. galaxies merged into one galaxy, these older galaxies are defined as the progenitors. There exists one particular galaxy along the merging history, which is the most massive one of all galaxies participate in the merging process. These most massive progenitors are defined as the main progenitors. We will mainly investigate the properties of galaxies and their progenitors.

%\rachel{Note: I fixed this here, but you should see if other places need correction - it should be `galaxy', not `Galaxy' (`Galaxy' is reserved for our Milky Way, while `galaxy' is for galaxies in general).}  
In this section, we describe the key steps involved in the construction of the galaxy merger tree. To begin with, halo/subhalo merger trees were identified by running the \texttt{ROCKSTAR}~\citep{behroozi2012rockstar} halo/subhalo finder along with \texttt{CONSISTENT-TREES}~\citep{behroozi2012gravitationally}, both of which are described in the following two subsections.
\subsubsection{\texttt{ROCKSTAR}}
\label{rockstar_sec}
\texttt{ROCKSTAR} (or `Robust Overdensity Calculation using K-Space Topologically Adaptive Refinement') is an algorithm based on adaptive hierarchical refinement of FOF groups. Primary FOF groups are first identified using a FOF finder. Within each FOF group, a hierarchy of FOF subgroups~(in phase space) is identified using an adaptive refinement of the linking length. The FOF subgroups at the lowest~(deepest) level of the hierarchy are then converted into seed haloes. Starting with the lowest level of the hierarchy, the FOF subgroup particles are assigned to the seed haloes based on phase space distances; this process is repeated for the higher levels of the hierarachy until all particles of the parent FOF group have been assigned to the halo. After assigning all the particles, the host-subhalo relationship is calculated by assigning a seed halo to be a \textit{subhalo} of the closest seed halo~(within the same FOF group) with larger number of assigned particles. This process is performed until all the seed haloes are either \textit{host haloes} or \textit{subhaloes}. For further implementation details, see the original paper \citep{behroozi2012rockstar}.  

\subsubsection{\texttt{CONSISTENT-TREES}}
\label{trees_sec}
We build a merger tree for our \texttt{ROCKSTAR} haloes/subhaloes using \texttt{CONSISTENT-TREES} algorithm \citep{behroozi2012gravitationally}. \texttt{CONSISTENT-TREES} is an extension to traditional \textit{particle based}~(constructed by tracing trajectories of halo/subhalo particles across different time steps) tree building algorithms which can potentially compromise the \textit{continuity} of halo/subhalo properties across simulation time-steps, due to the issues listed in Section 2.2 of \cite{behroozi2012gravitationally}.

%\rachel{There is a lot of `halo/subhalo' which should be either  `halo/subhalo' or `halo / subhalo'.  I fixed a bunch of them, but please go through consistently.  Also, I am not sure why there are so many italicized words  throughout the paper\dots it looks odd.}

\texttt{CONSISTENT-TREES} resolves the foregoing problem by tracing~(in addition to particles) a subset of halo/subhalo properties which include halo mass, maximum circular velocity, halo position, and bulk velocity. A major component of the algorithm is to ensure continuity in these halo properties by construction. This is achieved by running a particle-based tree finder and establishing preliminary links between progenitor haloes~(at time step $t_{\mathrm{n-1}}$) and descendant haloes~(at time step $t_{\mathrm{n}}$). The subsequent steps consist  of the following actions:
\begin{enumerate}
\item Gravitationally tracing the positions of descendant haloes from $t_{\mathrm{n}}$ to $t_{\mathrm{n-1}}$ to obtain their most likely progenitors at $t_{\mathrm{n-1}}$; removing progenitors whose properties do not resemble the most likely progenitors of the corresponding descendants.

\item For each descendant halo at $t_{\mathrm{n}}$ that lacks a progenitor at $t_{\mathrm{n-1}}$ after step (i), a \textit{phantom} progenitor is assigned with halo properties identical to its most likely progenitor at $t_{\mathrm{n-1}}$; however, those descendant haloes that do not have progenitors for a sufficiently large sequence of time steps are removed. 

\item Finally, if a halo at $t_{\mathrm{n-1}}$ has no descendant at $t_{\mathrm{n}}$ after step (ii), it is \textit{merged} with a halo~(at $t_{\mathrm{n}}$) in its vicinity that has the strongest tidal field; additionally, the halo is removed as a statistical fluctuation if it is too far away from other haloes to experience any significant tidal field. 

\item Steps (i) to (iii) are iterated over the range of time steps (where each iteration corresponds a pair of time slices $t_{n-1}$ and $t_n$) from final time $t_f$ to initial time $t_i$. This establishes a lineage of haloes over the time range $t_i$ to $t_f$.    

\end{enumerate}
Readers who are interested in more details are encouraged to refer to Section 5 of  \cite{behroozi2012gravitationally}.
\subsubsection{Constructing galaxy merger tree: Matching \texttt{ROCKSTAR} and \texttt{SUBFIND}}
\label{matching_sec}

The subhalo merger trees obtained using \texttt{ROCKSTAR-CONSISTENT TREES} are dark matter only. In order to construct the galaxy merger tree for our \texttt{SUBFIND} galaxies, we must match the subhaloes on the \texttt{ROCKSTAR} merger tree to our \texttt{SUBFIND} galaxies. We perform the following steps for the matching: 
\begin{enumerate}
\item For a given \texttt{ROCKSTAR} subhalo (mass $M_h^{\mathrm{RS}}$) denoted by \texttt{SUBHALO-RS}, we select all \texttt{SUBFIND} subhalos~(with mass $M_h^{\mathrm{sub}}$) which satisfy $0.5\times M_h^{\mathrm{RS}}<M_h^{\mathrm{sub}}<2\times M_h^{\mathrm{RS}}$ and within a maximum distance of $5\times R^{\mathrm{RS}}_{\mathrm{vir}}$, where $R^{\mathrm{RS}}_{\mathrm{vir}}$ is the virial radius of the \texttt{ROCKSTAR} subhalo. We then choose the \texttt{SUBFIND} subhalo that is closest to the \texttt{ROCKSTAR} subhalo, denoted by \texttt{SUBHALO-RS-SUB}.
\item For the \texttt{SUBFIND} subhalo \texttt{SUBHALO-RS-SUB}, we select all \texttt{ROCKSTAR} subhalos~(with mass $M_h^{\mathrm{sub}}$) which satisfy $0.5\times M_h^{\mathrm{sub}}<M_h^{\mathrm{RS}}<2\times M_h^{\mathrm{sub}}$ and within a maximum distance of $5\times R^{\mathrm{sub}}_{\mathrm{vir}}$, where $R^{\mathrm{sub}}_{\mathrm{vir}}$ is the virial radius of the \texttt{SUBFIND} subhalo. We then choose the \texttt{ROCKSTAR} subhalo that is closest to the \texttt{SUBFIND} subhalo, denoted by \texttt{SUBHALO-RS-SUB-RS}
\item If~(and only if) we retrieve the original \texttt{ROCKSTAR} subhalo at the end of step (ii), i.e., \texttt{SUBHALO-RS-SUB-RS} is identical to \texttt{SUBHALO-RS}, we say that \texttt{SUBHALO-RS}~(from the \texttt{ROCKSTAR} merger tree) and
\texttt{SUBHALO-RS-SUB}~(from the \texttt{SUBFIND} catalog) have been \textit{matched}.
\end{enumerate}
In order to generate a corresponding \texttt{SUBFIND} galaxy merger tree from a \texttt{ROCKSTAR} merger tree, every \texttt{ROCKSTAR} subhalo on the tree must be matched with a \texttt{SUBFIND} galaxy for the redshift range of our interest~($z_i\leq z\leq z_f$). If the matching fails at any redshift within ($z_i\leq z\leq z_f$), the entire tree is discarded. We quantify the matching success rate by defining a \textit{matching efficiency} $\eta_{\mathrm{matching}}$ as the ratio of the number of matched \texttt{SUBFIND} trees over the number of original \texttt{ROCKSTAR} trees~(present before matching). Figure \ref{matching_efficiency} shows $\eta_{\mathrm{matching}}$ as a function of $M_h$ at various values of $z_f$~($z_i=0.6$). %\rachel{You seem to switch between ($z_i, z_f$) and $(z_i, z_f)$ - what distinction is being made?} 
For $z_f=1.5$~(red line), the efficiency is $86\%$ for all masses. At higher $z_f$, we lose more trees~(as expected) and the efficiency decreases to $75-82\%$ for $z_f=3$. This translates to a total of 27942 \texttt{SUBFIND} galaxy merger trees with progenitors up to redshift 3. This sample is sufficient for a statistical analysis, and to avoid further decrease in efficiency, we choose not to trace progenitors beyond redshift 3,  hereafter defining the redshift range of our study to be $0.6\leq z\leq3$. We chose $z\geq 0.6$ since it is the time period when galaxy formation and merger processes are most active.

\subsection{Shapes of galaxies and dark matter halos}

\begin{figure*}
\begin{center}
 \includegraphics[width=1.17\textwidth]{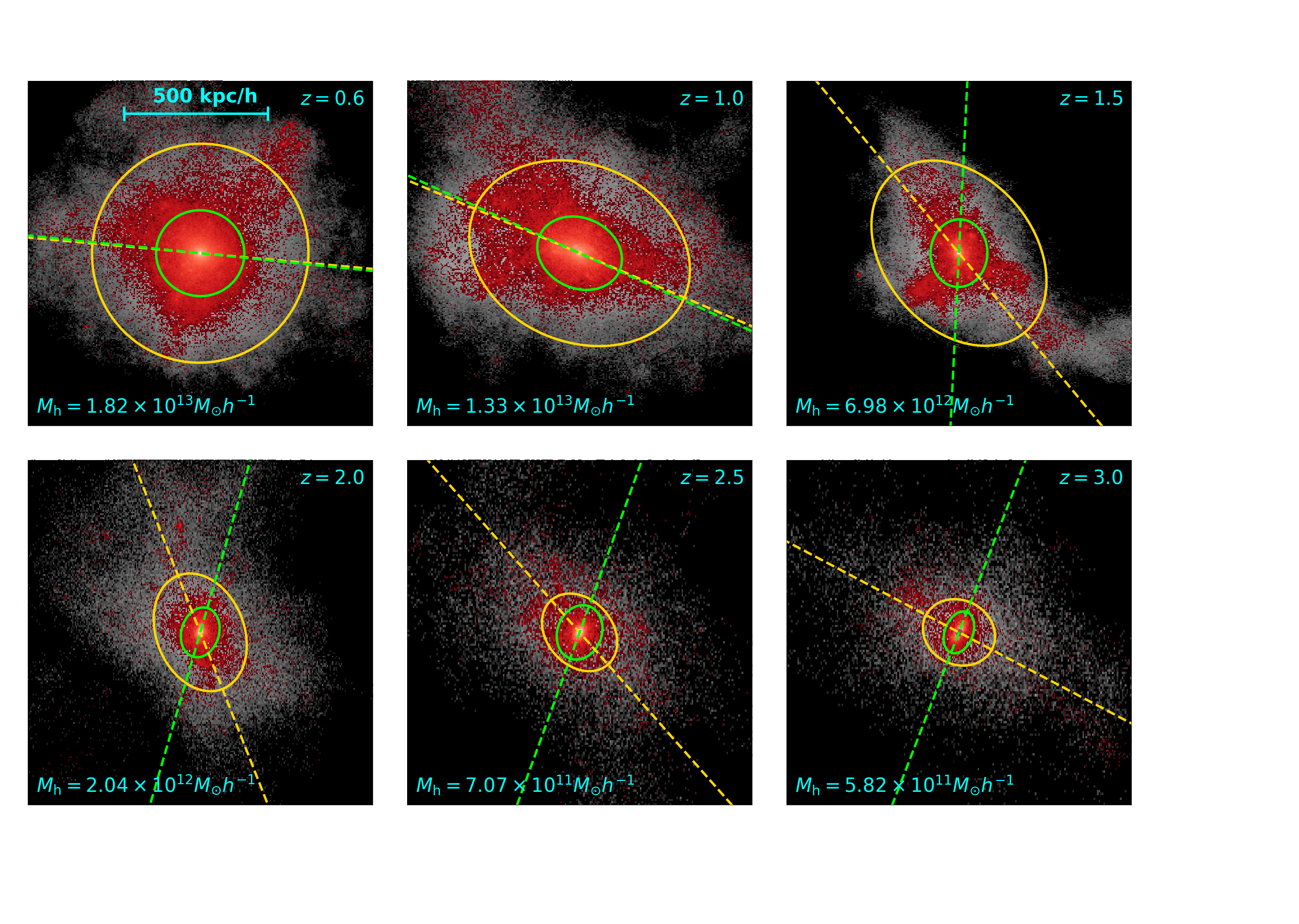}
\end{center}
 \vspace{-1cm}
 \caption{A 2-d illustrative example of the evolution of a MBII galaxy on the merger tree. The red histograms show the distribution of stars and grey histograms show the distribution of underlying dark matter. The yellow ellipse represents the shape identified using dark matter particles, while the green ellipse represents the shape identified using stellar matter particles; The yellow and green dashed lines of color are their corresponding major axis directions. We can see that the subhalo shape is becoming more spherical from $z=3$ to $z=0.6$. Furthermore, the alignment between stellar matter and dark matter shapes is becoming stronger as we go from $z=3$ to $z=0.6$.}
 \label{illustration}
\end{figure*}
%\rachel{Caption should give some lesson learned from this figure.  Also, it should say why the galaxy appears to wander around: how was the center position in each redshift chosen?

\begin{figure*}
 \includegraphics[width=130mm]{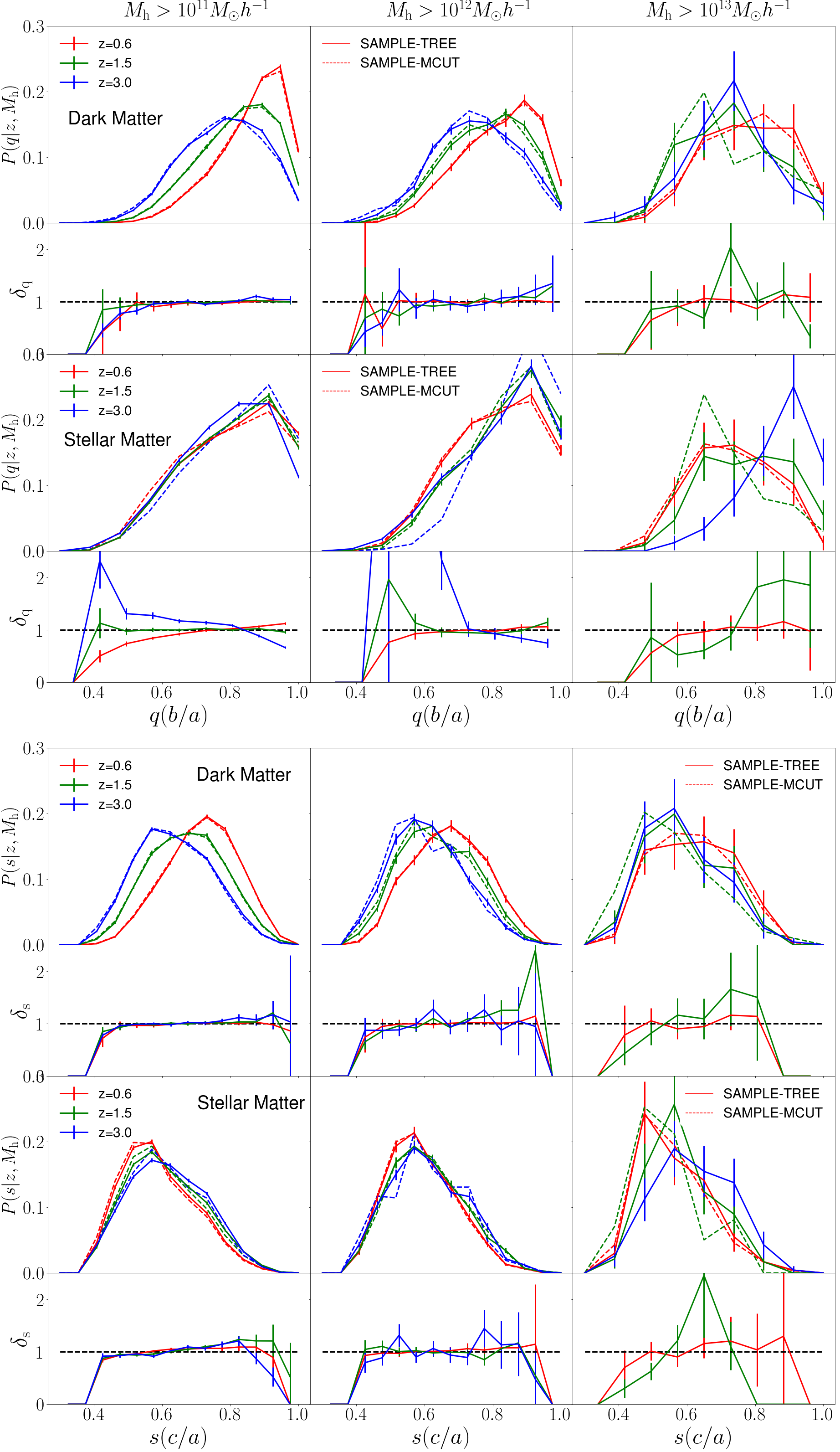}
 \caption{Distribution of galaxy shapes: $P(q|z,M_h)$ (top) and $P(s|z,M_h)$ (bottom) show the normalized probability distributions of axis ratios $q=\frac{b}{a}$ and $s=\frac{c}{a}$ of dark/stellar matter components of galaxies~(subhaloes). Solid lines and dashed lines correspond to galaxy samples \texttt{SAMPLE-TREE} and \texttt{SAMPLE-MCUT} respectively~(see section \ref{sample_definitions} for definition of galaxy samples). $\delta_q$ and $\delta_s$ correspond to the ratio of $P(q|z,M_h)$ and $P(s|z,M_h)$ respectively between \texttt{SAMPLE-TREE} and \texttt{SAMPLE-MCUT} galaxies. The errorbars are $1\sigma$ poisson errors.} 
 \label{fig:q}
\end{figure*}

We now describe how galaxy shapes are quantified. We model the shapes of the dark matter and stellar matter components of subhalos as ellipsoids in three dimensions by using the eigenvalues and eigenvectors of the \textit{reduced} inertia tensor~\citep{2005ApJ...627..647B,tenneti2014galaxy} given by 
\begin{equation}
I_{ij}=\frac{\Sigma_{n}m_{n}\frac{x_{ni}x_{nj}}{r_n^2}}{\Sigma_{n}m_{n}}
\label{inertia_tensor}
\end{equation}
where $m_n$ is  the mass of the $n^{th}$ particle and $x_{ni}$ and $x_{nj}$ represent the $i$ and $j$ component of the position of the $n^{\rm th}$ particle ($0\le i,j\le 2$). $r_n$ is the distance of the $n^{\rm th}$ particle from the subhalo center and is given by $r_n^2=\sum x_{ni}^2$.

We denote the principal axis directions or eigenvectors~(unit vectors) of $I_{ij}$ to be $(\hat{e}_a, \hat{e}_b, \hat{e}_c)$ with corresponding eigenvalues $(\lambda_a, \lambda_b, \lambda_c)$. The lengths of the principal axes $(a,b,c)$ are given by $(\sqrt{\lambda_a}, \sqrt{\lambda_b}, \sqrt{\lambda_c})$. The ellipticities can then be measured by the axis ratios,
\begin{equation}
q=\frac{b}{a},s=\frac{c}{a}.
\label{axes_ratio}
\end{equation}
where $a$ is the length of the primary~(largest) axis. A perfectly spherical subhalo corresponds to $q=s=1$ and a triaxial halo corresponds to $q\neq s<1$.

%All particles are given equal weight. Alternatively, one can also define a \textit{reduced} inertia tensor, which gives more weight to particles which are closer to the center, and is given by
%\begin{equation}
%I_{ij}=\frac{\sum_{n}\frac{m_{n}x_{ni}x_{nj}}{r^2_n}}{\sum_{n}m_{n}}
%\end{equation}
%where $r^2_n= \sum_i x^2_{ni}$.

For a more robust measure of the shape, we adopt an iterative approach wherein we first determine the principal axes and axis ratios using all the particles in the subhalo, thereby determining the ellipsoidal volume. For each successive iteration, we then recalculate the inertia tensor and axis ratios ignoring particles outside the ellipsoidal volume. We repeat this until each iteration leads to $\lesssim1\%$ change in $a$, $b$ and $c$.  

\subsubsection{Shape convergence test} \label{S:appendixa}
We require a sufficiently large number of particles to reliably measure galaxy~(subhalo) shapes. Here, we determine the minimum number of particles. Figure~\ref{shape_convergence} shows the distribution of $q$~(denoted by $P(q|M_h)$) for $z=0.6$ and $M_h>10^{11}~M_{\odot}/h$ galaxies. We show $P(q|M_h)$ for different numbers~($N_{\mathrm{part}}$) of subsampled dark matter particles within each subhalo. We find that the distributions converge for $N_{\mathrm{part}}=300,500,1000$ whereas for $N_{\mathrm{part}}=50$, $q$ is significantly underestimated. Therefore, we assume $N_{\mathrm{part}}\geq300$ in this work to ensure shape convergence; similarly, this choice is also sufficient for the convergence of $s$. This sets a minimum subhalo mass of our galaxies to $M_h\sim 3\times10^{9}~M_{\odot}/h$, which limits the subhalo mass and redshift range over which we can construct merger trees. We find that for galaxies with $M_h>10^{11}~M_{\odot}/h$ at $z=0.6$, their progenitors have $M_h\gtrsim3\times10^{9}~M_{\odot}/h$ up to $z=3$. Therefore our final choice for the subhalo mass range and redshift range in this work are $M_h>10^{11}~M_{\odot}/h$ and $0.6<z<3$.

%\rachel{Cite a reference for this definition of inertia tensor.}
%\begin{equation}
%r^2_n= \frac{x^2_{n0}}{a^2} + \frac{x^2_{n1}}{b^2} + %\frac{x^2_{n2}}{c^2}
%\end{equation}
%where $a$, $b$, $c$ are half-lengths of the principal axes of the ellipsoid and are all equal to $1$ in the first iteration.
%This method corresponds more closely to observational shape measurements such as the ones based on weighted quadrupole moments \citep{kirk2015galaxy}.
%Fig. 4 shows the histogram of the axis ratio $q=\frac{b}{a}$ of dark matter subhalo with mass $>10^{11}M_{\odot}h^{-1}$ on the merger tree at $z=0.6$ distribution under different shape definitions. From Fig. 4, we can see that the unweighted axis ratio distributions obtained with non-iterative and iterative  definitions are identical. For the reduced inertia tensor, the results for the iterative calculation are uniformly more flattened than for the non-iterative calculation. The reason for this is that the non-iterative reduced calculation implicitly imposes spherical symmetry (via the $1/r^2$ weighting)\aklant{I don't understand the "spherical symmetry" argument. I think I mentioned this before but you did not change it much. Please talk to Rachel/ Tiziana to make sure this part makes sense. Maybe I'm missing something}, which will result in an overly-spherical shape estimate. For this reason, we will only take unweighted and reduced iterative definition of shapes.
\subsection{Misalignment angle}
To quantify the misalignment between the galaxy~(stellar matter component) and its host dark matter subhalo, we calculate the principal axes corresponding to the dark matter and star particles, i.e., $(\hat{e}^{\mathrm{DM}}_a, \hat{e}^{\mathrm{DM}}_b, \hat{e}^{\mathrm{DM}}_c)$ and  $(\hat{e}^{*}_a, \hat{e}^{*}_b, \hat{e}^{*}_c)$ respectively. The misalignment angle is then defined as the angle between the eigenvectors corresponding to the primary~(longest) axes. 
\begin{equation}
\theta_{m}=\arccos\left(\left| \hat{e}^{\mathrm{DM}}_{a} \cdot \hat{e}^{*}_{a}\right| \right)
\end{equation}

\subsection{Correlation function}
The ellipticity-direction~(ED) correlation function \citep{lee2008quantifying} %{mandelbaum2006detection}
 cross-correlates the orientation of the major axis of a subhalo with respect to the large-scale density field. For a subhalo centered at position $x$ with major axis direction $\hat{e}_{a}$, the ED cross-correlation function is given by
\begin{equation}
\omega \left(r\right) = \left \langle \left| \hat{e}_{a}(\vec{x}) \cdot  \hat{r}(\vec{x}+\vec{r}) \right|^2 \right \rangle -\frac{1}{3}
\end{equation}
where $\hat{r}=\frac{\vec{r}}{r}$ and $\vec{r}$ is the position vector originating from the subhalo position~($\vec{x}$) to a tracer~(galaxy positions or dark matter particle positions) of the large scale matter distribution around the halo. In this work, we have used the dark matter particle positions as tracers of the matter density field.

\begin{figure}
 \includegraphics[width=0.5\textwidth]{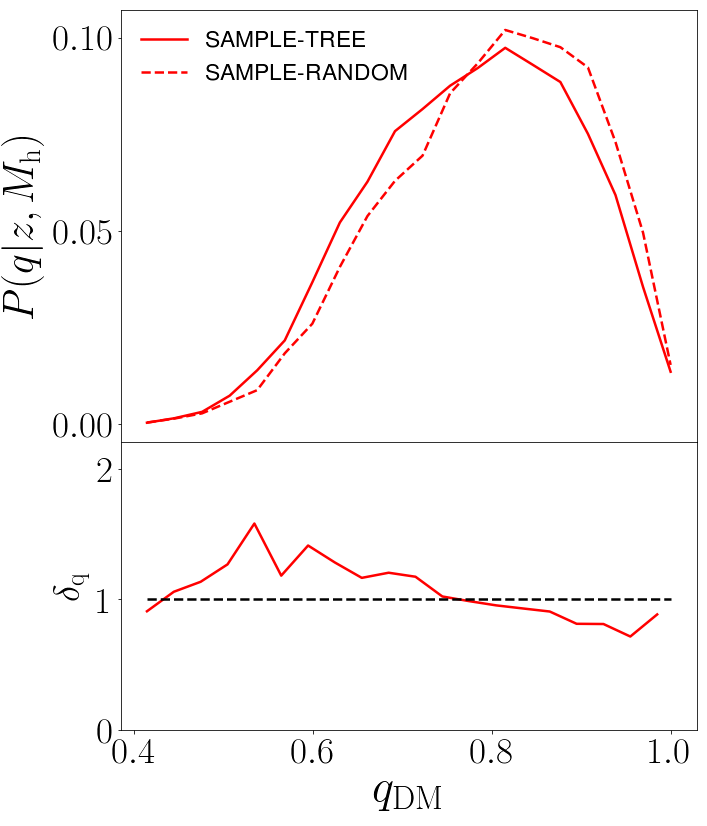}
 \caption{Comparison of the shapes of \textit{progenitor} galaxies and \textit{randomly-selected} galaxies of similar mass at $z=z_f=3$. The solid and dashed lines show $P(q|z,M_h)$~(dark matter component) for \texttt{SAMPLE-TREE}: $M_h>10^{11}~M_{\odot}/h$: $z=3$ and \texttt{SAMPLE-RANDOM}: $M_h>10^{11}~M_{\odot}/h$: $z=3$~(see Section~\ref{sample_definitions} for the sample definitions). $\delta_q$ is the ratio between the solid and dashed lines. \texttt{SAMPLE-RANDOM} is constructed to have a mass function identical to that of \texttt{SAMPLE-TREE} progenitors. }
 \label{fig:random}
\end{figure}

\begin{figure*}
 \includegraphics[width=\textwidth]{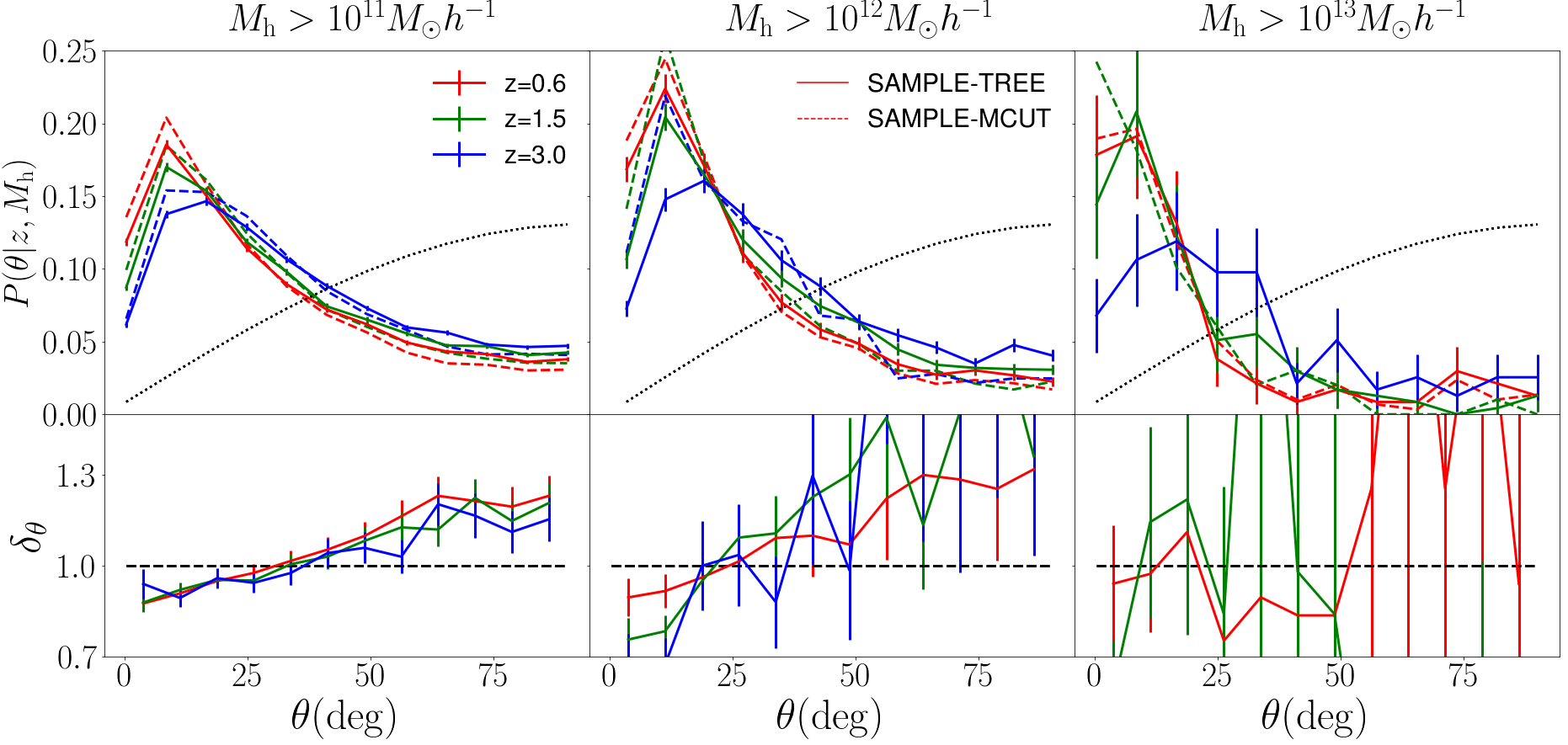}
 \caption{$P(\theta|z,M_h)$ is the distribution of misalignment angle $\theta$ between stellar and dark matter matter component of subhalos. Solid and dashed lines correspond to \texttt{SAMPLE-TREE} and \texttt{SAMPLE-MCUT} galaxies, respectively. $\delta_\theta$ is the ratio between the solid and dashed lines. The errorbars are $1\sigma$ poisson errors. The black dotted lines represent the misalignment angle distribution if the two eigenvectors are uniformly distributed in 3d space.}
 \label{fig:theta_hist}
\end{figure*}

\section{Results} \label{S:results}

\subsection{Stellar mass-subhalo mass relation}

Figure \ref{SM_HM_fig} shows the subhalo total~(dark matter+gas+stars+black hole) mass~($M_h$) versus stellar mass~($M_*$) relation of \texttt{SAMPLE-TREE} galaxies at $z=0.6$, $1.5$, and $3.0$ with  $M_h>10^{11}M_{\odot}/h$ at $z=0.6$. As expected, $M_h$ and $M_*$ are strongly correlated and both decrease with increasing redshift. We also note that as redshift increases, the $M_h$-$M_*$ relation does not significantly change either in slope or intercept, broadly consistent with predictions from semi-analytical models~\citep{2016MNRAS.456.1459M} as well as observations \citep{2012ApJ...744..159L}. This implies that galaxies grow in stellar mass and dark matter mass at roughly the same rate as they evolve along the merger tree.%\rachel{This has been studied quite a bit, observationally and theoretically, so it would be good to comment on how this compares with other work.  For example, \url{https://arxiv.org/pdf/1104.0928.pdf} figures 11+.}

As the subhalo mass strongly correlates with stellar mass, and therefore also correlates with other observable properties such as luminosity, star formation rate, we shall hereafter use subhalo mass cuts to construct the various galaxy samples defined in the next section for the rest of this work. 

\subsection{List of galaxy samples: Definitions and notations}
\label{sample_definitions}
Before we discuss the rest of the results, we describe the types of galaxy samples that we consider in this work.
\begin{itemize}
\item \texttt{SAMPLE-TREE}: The primary sample of interest consists of galaxies on the merger tree. We select galaxies with different subhhalo mass cuts~($M_h$) at $z=0.6$ and trace their progenitors to $z=3$ using the methods described in Sections \ref{merger_tree_sec}. Hereafter, we shall refer to this sample as \texttt{SAMPLE-TREE}. For example, the sample name  ``\texttt{SAMPLE-TREE}: $M_h>10^{11}~M_{\odot}/h$: $z=2$" refers to galaxies at $z=2$ that are progenitors of the $M_h>10^{11}~M_{\odot}/h$ galaxies as selected at $z=0.6$. Using this sample, we study the redshift evolution of IA properties of galaxies, without having to consider the impact of evolution due to sample selection.
\item \texttt{SAMPLE-MCUT}: The secondary sample of interest is obtained using the selection criterion of \cite{tenneti2015intrinsic}. Here we select galaxy samples with a fixed subhalo mass cut applied at all redshifts. Hereafter, we shall refer to this sample as \texttt{SAMPLE-MCUT}. For example, the sample name ``\texttt{SAMPLE-MCUT}:
$M_h>10^{11}~M_{\odot}/h$: $z=2$" refers to all galaxies at $z=2$ with $M_h>10^{11}~M_{\odot}/h$. With this sample, the observed redshift evolution of IA properties is a combination of \textit{intrinsic} redshift evolution effects, and the evolution due to sample selection.
\item \texttt{SAMPLE-RANDOM}: To interpret the impact of requiring galaxies to be a part of a merger tree, it will be necessary to look at differences in IA properties between a progenitor~(merger tree) galaxy and a randomly chosen galaxy of similar mass. To do this, we construct a galaxy sample by randomly drawing galaxies from the full sample at some redshift~(all galaxies in the simulation snapshot), such that the total~(dark matter+gas+stars+black hole) mass function is modulated to be identical to that of \texttt{SAMPLE-TREE}~(progenitor) galaxies at the same redshift. We shall refer to this as sample \texttt{SAMPLE-RANDOM}. For example, the sample name  ``\texttt{SAMPLE-RANDOM}: $M_h>10^{11}~M_{\odot}/h$: $z=2$" refers to a random sample of galaxies at $z=2$ whose mass function is identical~(by construction) to ``\texttt{SAMPLE-TREE}: $M_h>10^{11}~M_{\odot}/h$: $z=2$".

\end{itemize}
\subsection{Evolution of galaxy shapes and misalignment angles}
In this subsection, we will investigate how the shapes of galaxies~(and dark matter subhaloes), described by axis ratios $q=\frac{b}{a}$ and $s=\frac{c}{a}$, and the misalignments between stellar and dark matter components, evolve with redshift along the merger tree. Figure~\ref{illustration} shows an illustration of the evolution of a single simulated galaxy along the merger tree from $z=3$ to $z=0.6$. We can see that the shape of the dark matter component~(yellow ellipse) becomes more spherical with decreasing redshift. Furthermore, at $z=3$, the stellar matter is significantly misaligned with respect to the dark matter, but the alignment becomes stronger as redshift decreases. In the following subsections, we shall show that the foregoing trends persist for the overall distribution of shapes and misalignment angles for the entire set of \texttt{SAMPLE-TREE} galaxies.  
   
\subsubsection{Shape}

\begin{figure*}
 \includegraphics[width=\textwidth]{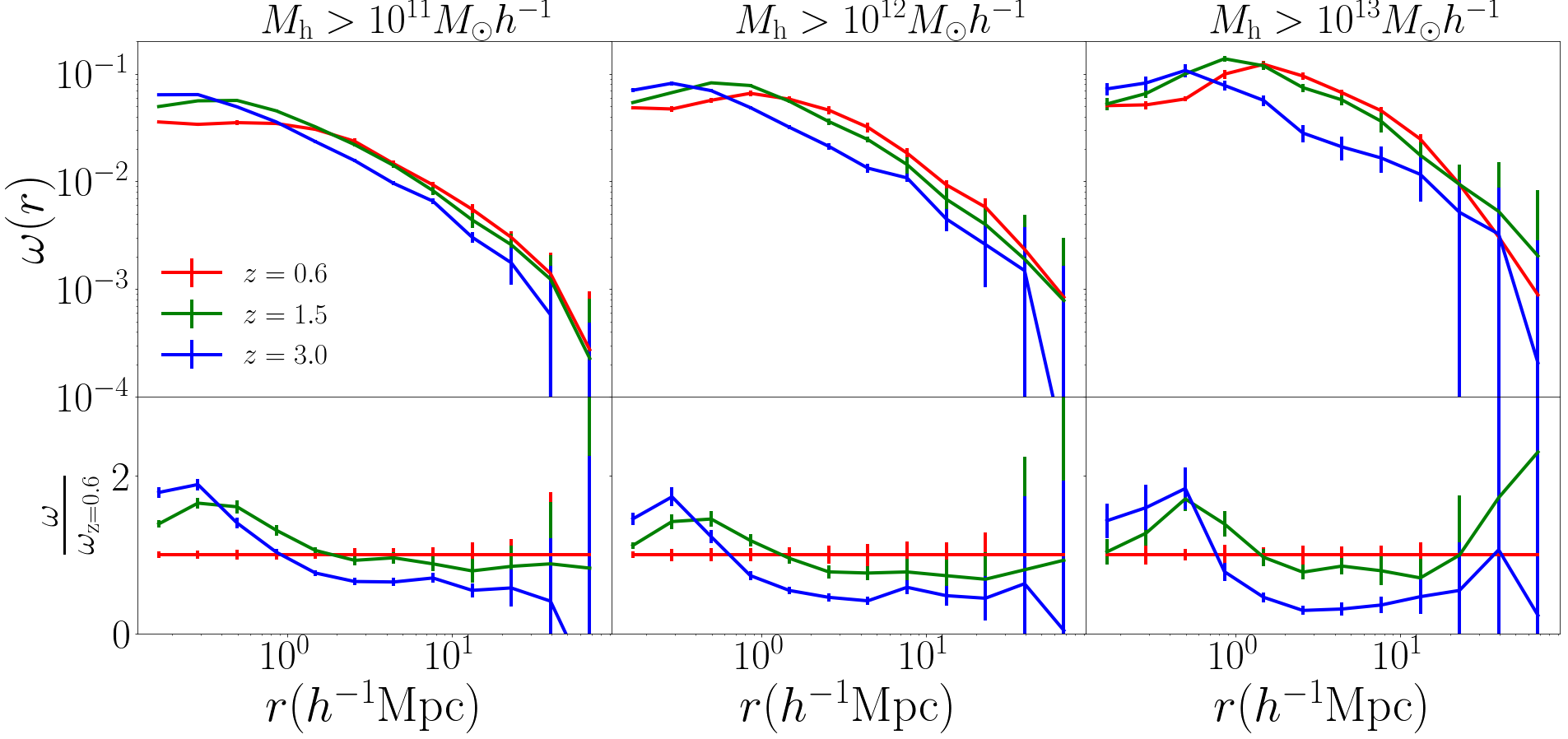}
 \caption{$\omega(r)$ is the ellipticity direction~(ED) correlation function of \texttt{SAMPLE-TREE} galaxies at different redshifts. Here we are using the major axes of the stellar matter components and galaxy positions as tracers of the matter distribution. The bottom panels show the ratio of $\omega(r,z)$ with respect to that of $\omega(r,z=0.6)$.  Errorbars are jackknife errors obtained by dividing the simulation volume into eight octants.} %\rachel{Is that how all errorbars on previous plots were obtained? It seems odd to only mention it now. If you use jackknife consistently throughout the paper, then you might make a short subsection on this in the methods section and just talk about it there.}}
 \label{fig:corr}

 \includegraphics[width=\textwidth]{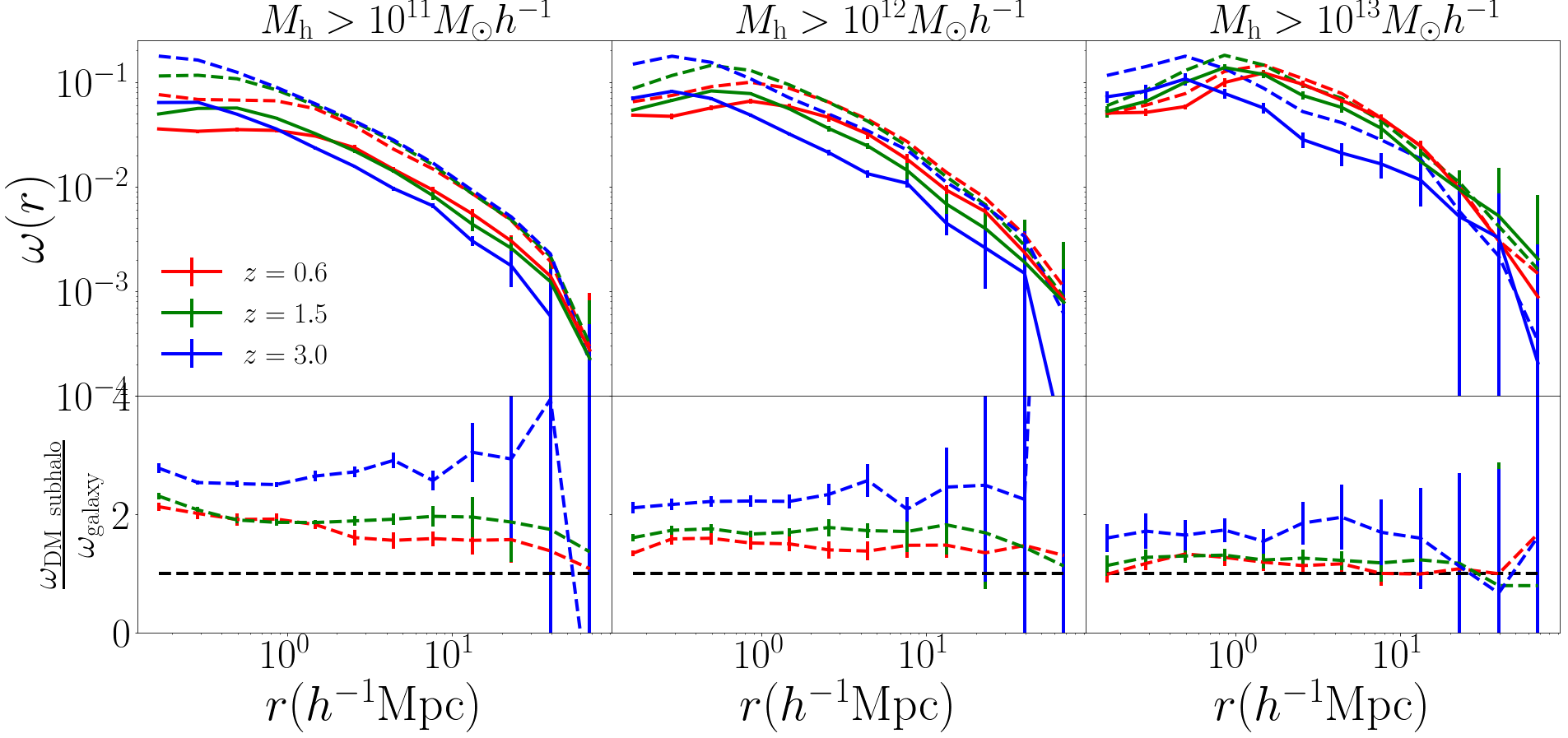}
 \caption{\textbf{Comparing ED correlation functions for \texttt{SAMPLE-TREE} galaxies and their dark matter subhaloes:} In the top panels, solid and dashed lines show the ED correlation functions of galaxies and their dark matter subhaloes, respectively. The ratio between the dashed vs solid lines are shown in the bottom panels. Errorbars in the correlation function are jackknife errors obtained by dividing the simulation volume into eight octants.}
 %\rachel{I don't think this can be right, because solid is clearly below dashed, yet the bottom panel exceeds 1.}
 \label{fig:dm_bm}
\end{figure*}

\begin{figure*}
 \includegraphics[width=\textwidth]{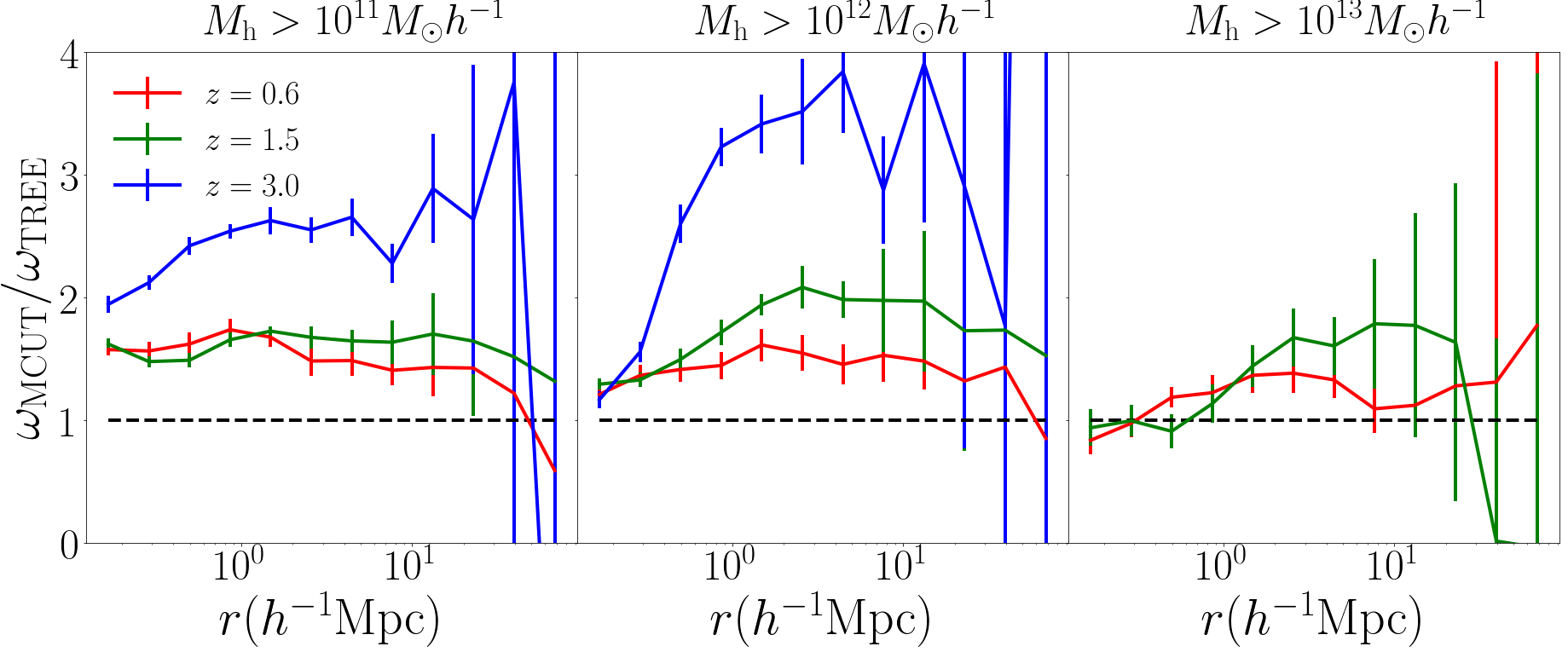}
 \caption{$\omega_{\texttt{MCUT}}/\omega_{\texttt{TREE}}$ is the ratio of $\omega(r)$ of \texttt{SAMPLE-TREE} to that of \texttt{SAMPLE-MCUT} galaxies. Errorbars are jackknife errors obtained by dividing the simulation volume into eight octants.}
 \label{fig:comparison}
\end{figure*}

Figure~\ref{fig:q} shows the distributions $P(q|z,M_h)$ and $P(s|z,M_h)$ of axis ratios $q$ and $s$ respectively. In Section \ref{S:appendixa}, we established that $\gtrsim$300 particles are required to reliably measure the shape; this dictates our choice of minimum subhalo mass threshold of $M_h>10^{11}~M_{\odot}/h$ at $z=0.6$. %We do not consider galaxies and their progenitors with masses below $M_h>10^{11}~M_{\odot}/h$ at $z=0.6$.%\rachel{That entire caveat `In Appendix\dots at $z=0.6$' probably better belongs in section 2.4, where one might naturally ask how many particles is needed to achieve convergence.} 
The solid and dashed lines correspond to \texttt{SAMPLE-TREE} and \texttt{SAMPLE-MCUT} respectively. The bottom panels show the ratio between the axis ratio distributions of \texttt{SAMPLE-TREE} and \texttt{SAMPLE-MCUT} galaxies.

\textbf{Subhalo mass dependence on the merger tree}:
We first focus on shapes of dark matter subhaloes. For \texttt{SAMPLE-TREE} galaxies~(solid lines), we see that as subhalo mass increases, $P(q|z,M_h)$ and $P(s|z,M_h)$~(for dark matter) is increasingly skewed towards lower values of $q$ and $s$. This is more clearly seen in the mean values of $q$ and $s$ in Figures B1. This implies that as subhalo mass increases, galaxies on the merger tree become less spherical at fixed redshift. This is also true for \texttt{SAMPLE-MCUT} galaxies~(dashed lines) and has been well established in previous studies~\citep{2005ApJ...618....1H,2006MNRAS.367.1781A,tenneti2015intrinsic}; therefore it is not surprising that it persists for galaxies on the merger tree.

For the shapes of the stellar matter component, the dependence on subhalo mass at $z\lesssim1.5$ is the same as that of the dark matter component for both $P(q|M_h)$ and $P(s|M_h)$, also seen in \cite{tenneti2015intrinsic}. In other words, at $z\lesssim1.5$ more massive galaxies have less spherical shapes for the stellar matter component~(the mass dependence is seen much more clearly in Figure~B2). However, this result does not persist all the way up to $z\sim3$. In fact we see that the mass dependence of $P(q|M_h)$ is reversed~(i.e.\ $P(q|M_h)$ skews towards higher values with increasing subhalo mass) at $z\sim3$ while $P(s|M_h)$ has no significant mass dependence at $z\sim3$. Therefore, we find that at $z\sim3$, the sphericity of the stellar matter component of galaxies increases with increasing subhalo mass.

To summarize the above trends, we find that:
\begin{itemize}
    \item The shapes of the dark matter components of galaxies become less spherical with increasing subhalo mass.
    \item For the stellar matter components, the shapes become less spherical with increasing subhalo mass at $z\lesssim1.5$. The trend starts to reverse at $z\gtrsim1.5$ and by $z\sim3$, the shapes become more spherical with increasing subhalo mass. 
    %As we shall see in the following paragraphs, the reversal is due to the subhalo mass dependence in the redshift evolution of the stellar matter shapes.}  
\end{itemize}

\textbf{Redshift evolution on the merger tree}:
We first focus on the shapes of dark matter subhaloes. For \texttt{SAMPLE-TREE} galaxies~(solid lines), we see that for all three panels, as redshift decreases, the peaks of $P(q|z,M_h)$ and $P(s|z,M_h)$~(for dark matter) shift towards higher values of $q$ and $s$. This implies that as redshift decreases, galaxies on the merger tree evolve to become more spherical. This is also true for \texttt{SAMPLE-MCUT} galaxies~(dashed lines), as was previously reported in \cite{tenneti2015intrinsic}. It also noteworthy that our results are consistent with \cite{2005ApJ...618....1H} which investigated the evolution of shapes of cluster sized haloes~($M_h>2\times10^{13}~M_{\odot}/h$) in N-body simulations over roughly the same range of redshifts. 

The shape evolution of the stellar matter component has significant differences compared to that of dark matter (as already hinted in the discussion on the subhalo mass dependence). For instance, $P(s|z,M_h)$ tends towards being less spherical as redshift decreases. This trend is opposite to that of dark matter. However, note also that the overall evolution of $P(s|z,M_h)$ is significantly weaker for stellar matter than for dark matter. For $P(q|z,M_h)$, the evolution is more complicated and depends on the subhalo mass threshold. For $M_h>10^{11}~M_{\odot}/h$, there is no significant evolution. On the other hand, for $M_h>10^{12}~M_{\odot}/h$ and $M_h>10^{13}~M_{\odot}/h$, the evolution is significant: $P(q|z,M_h)$ is less spherical at $z=0.6$ compared to $z=3$.

To summarize the above trends, we find that: 
\begin{itemize}
\item The shapes of the dark matter components of galaxies tend to become more spherical with time.

\item The shapes of the stellar matter components of galaxies tend to become less spherical with time, especially for higher mass thresholds.
\end{itemize}

\textbf{Comparing \texttt{SAMPLE-TREE} and \texttt{SAMPLE-MCUT}}:
We now compare the axis ratio distributions between \texttt{SAMPLE-TREE} and \texttt{SAMPLE-MCUT}~(see ratio plots in Figure \ref{fig:q}). 

For the dark matter shapes, we find that the axis ratio distributions of \texttt{SAMPLE-TREE} and \texttt{SAMPLE-MCUT} are broadly consistent i.e.\ there is no statistically significant difference in their shapes given the errorbars.  The fact that this is persistent all the way up to $z=3$ is noteworthy because at $z=3$, \texttt{SAMPLE-MCUT} galaxies are significantly more massive than \texttt{SAMPLE-TREE} galaxies. This suggests that at fixed redshift, the subhalo mass is not the sole parameter that determines the shapes of dark matter component of galaxies. In particular, galaxies that are progenitors of lower redshift galaxies above some mass threshold may be less spherical compared to a randomly chosen set of galaxies of similar subhalo mass. In order to show this explicitly, in Figure~\ref{fig:random} we compare the axis ratio distributions~(at $z=3$) of the dark matter components of \texttt{SAMPLE-TREE} galaxies to that of a random sample~(\texttt{SAMPLE-RANDOM}) whose mass functions are modulated to be identical to that of \texttt{SAMPLE-TREE}. We see that the axis ratios for \texttt{SAMPLE-TREE} galaxies are smaller than that of \texttt{SAMPLE-RANDOM} galaxies. This is also true in general for $z\gtrsim1.5$. This solidifies the impression that early galaxies that are progenitors of present-day massive galaxies~($M_h>10^{11}~M_{\odot}/h$ at $z=0.6$) are more elliptical~(on an average) than a randomly selected galaxy at similar subhalo mass and redshift.

For the stellar matter shapes, the ratio plots show that at $z=3$, $P(q|M_h)$ for samples with mass thresholds of $M_h>10^{11}~M_{\odot}/h$ and $M_g>10^{12}~M_{\odot}/h$ are less spherical for \texttt{SAMPLE-TREE} galaxies compared to \texttt{SAMPLE-MCUT} galaxies. This is because \texttt{SAMPLE-MCUT} galaxies are more massive compared to \texttt{SAMPLE-TREE} galaxies at $z=3$ (we have already shown that stellar matter shapes are more spherical at higher subhalo masses at $z=3$). $P(s|M_h)$ however has no significant difference between \texttt{SAMPLE-TREE} and \texttt{SAMPLE-MCUT} at $z=3$ despite the difference in subhalo masses. This is simply because there is insignificant mass dependence in $P(s/M_h)$ for stellar matter at $z=3$.

The comparison of shapes between \texttt{SAMPLE-TREE} and \texttt{SAMPLE-MCUT} galaxies at $z=3$ can now be summarized as follows:
\begin{itemize}
    \item For the dark matter components, no difference is found between the shapes of \texttt{SAMPLE-TREE} and \texttt{SAMPLE-MCUT} galaxies at $z=3$ despite the difference in masses. This is because at $z=3$ galaxies that are progenitors of $z\sim0.6$: $M_h\gtrsim10^{11}~M_{\odot}/h$ galaxies are significantly less spherical~(on an average) than a \textit{randomly selected} galaxy of similar subhalo mass and redshift.
    
    \item For the stellar matter component, \texttt{SAMPLE-TREE} galaxies are less spherical compared to \texttt{SAMPLE-MCUT} galaxies at $z=3$. This is because \texttt{SAMPLE-MCUT} galaxies are more massive (which we show to be more spherical for stellar matter component) than \texttt{SAMPLE-TREE} galaxies at $z=3$.

\end{itemize}

\subsubsection{Misalignment angle}

In this section, we investigate how the misalignment angle of galaxies on the tree evolves with redshift.

The solid lines in Figure~\ref{fig:theta_hist}~(top panels) show the distribution~$P(\theta|z,M_h)$ of misalignment angles~($\theta$) at different redshifts and subhalo mass cuts for \texttt{SAMPLE-TREE} galaxies. The distributions are skewed with a maximum at $\theta_m\sim5-10~\mathrm{deg}$ accompanied by a long tail at $\theta_m>10~\mathrm{deg}$, and a sharp fall-off at $\theta_m<5~\mathrm{deg}$. At fixed redshift, as the subhalo mass increases, $P(\theta|M_h)$ skews towards smaller values of  $\theta$~(seen more clearly in Figure B3). This implies that more massive galaxies are more aligned with their subhaloes. $P(\theta|M_h)$ skews towards smaller $\theta$ as redshift decreases, implying that galaxies evolve over time to become increasingly aligned with their subhaloes, although the evolution is mild.
%In addition, the tail of distribution also becomes smaller with decreasing redshift, thereby decreasing the fraction of galaxies with $\theta_m>10~\mathrm{deg}$  at decreasing redshifts. This suggests that galaxies tend to become more aligned with respect to their host halo/subhalo as redshift decreases, although the evolution is mild. 

%\rachel{Somewhere, either in the caption or text, we need to clearly emphasize that {\em randomly misaligned galaxies do not have a flat $\theta$ distribution} since this is 3D!  We could consider plotting in terms of something that should be flat, or show a random distribution for comparison.}

The evolution of the misalignment angle can be put in the context of existing IA models. The fact that the evolution is mild suggests that it may possibly be mediated by the evolution of the instantaneous tidal field. This is hinted by the fact that the contribution of the instantaneous tidal field is small~(compared to observations), as predicted by the analytical model presented in \cite{2015A&A...575A.113C}. In such a scenario, the redshift evolution, contributed by the instantaneous tidal field, can be thought of as a perturbation to the pre-existing alignment~($\theta_m\sim10~\mathrm{deg}$). Given its strength, the pre-existing alignment is likely set by the primordial~(at the formation epoch of these galaxies) tidal field, as assumed in linear alignment models~\citep{2001MNRAS.320L...7C,hirata2004intrinsic}.  

%\aklant{The fact that galaxies are strongly aligned with their subhaloes at all redshifts implies that these alignments were imprinted by the local~(\textit{primordial}) tidal field at their formation, as assumed by linear alignment model~\citep{2001MNRAS.320L...7C,hirata2004intrinsic}.
  
We also compare $P(\theta|z,M_h)$ for \texttt{SAMPLE-TREE} galaxies to the predictions for \texttt{SAMPLE-MCUT} galaxies~(solid vs.\ dashed lines in Figure \ref{fig:theta_hist}~top panels); Figure \ref{fig:theta_hist}~(bottom panels) shows the ratio $\delta_{\theta}$. For $M_H>10^{11,12}~M_{\odot}/h$, we find that $\delta_{\theta}<1$ for $\theta<25~\mathrm{deg}$ and  $\delta_{\theta}>1$ for $\theta>25~\mathrm{deg}$ at all redshifts. This implies that \texttt{SAMPLE-TREE} galaxies are less aligned with their subhaloes compared to \texttt{SAMPLE-MCUT} galaxies. At $z=1.5$ and $z=3$, one would expect this to be the case as \texttt{SAMPLE-MCUT} galaxies are more massive, and therefore more aligned, than \texttt{SAMPLE-TREE} galaxies. However, we also see the same effect at $z=0.6$, where both \texttt{SAMPLE-MCUT} and \texttt{SAMPLE-TREE} galaxies have the same subhalo mass thresholds. This implies that galaxies which formed between $0.6\lesssim z\lesssim 3$~(i.e. those that do not have progenitors up to $z=3$) are more aligned with their subhaloes than those that formed at $z>3$.%\rachel{I'm confused, those lines  don't look flat: it looks like at low mass, $\delta_\theta$ is clearly $<1$ at small angles, $>1$ at large angles.  Shouldn't we discuss and interpret this?  Tree galaxies are less aligned?}

We have so far discussed the evolution of distributions of galaxy shapes and misalignment angles. In Appendix~\ref{S:appendixb}, we present the evolution of the average values of the axis ratios and misalignment angles, and provide simple fitting functions to quantify them.  

\subsection{Ellipticity-direction~(ED) Correlation function}

In this section, we will investigate how the ellipticity-direction correlation function of galaxies on the merger tree evolves with redshift
We now present the results for the ED correlation function $\omega(r)$. The top panels in Figure~\ref{fig:corr} show $\omega(r)$ for \texttt{SAMPLE-TREE} galaxies and its redshift evolution along the merger tree. The bottom panels show the ratio $\omega(r,z)/\omega(r,z=0.6)$. They reveal the evolution of the ED correlation for a wide range of scales to be probed by LSST weak lensing~\citep{2018arXiv180901669T}. These include scales $\gtrsim5~\mathrm{Mpc}/h$ where the NLA model and its extensions such as \cite{2016IAUS..308..452B} already work well. Additionally, our simulations also reveal ED correlations at smaller scales which are not well probed by these analytical models. Accordingly, we choose $\sim 1~\mathrm{Mpc}/h$ as an interesting scale around which we shall now describe the evolution of the ED correlation.

At $r>1~\mathrm{Mpc}/h$ we see that the correlation function is a power law as a function of $r$. The slope of the power law does not vary significantly with redshift or subhalo mass. The power law amplitude increases with subhalo mass at fixed redshift, as also reported in \cite{tenneti2015intrinsic}. The ED correlation amplitude increases with decreasing redshift along the merger tree~(up to factors of $\sim4$ from $z=3$ to $z=0.6$).

At sufficiently small scales~($r\lesssim1~\mathrm{Mpc}/h$), $\omega(r)$ deviates from a  power law and is suppressed~(compared to power-law extrapolation from large scales). The extent of the suppression increases with decreasing redshift. As we approach even smaller scales $\sim 0.1~\mathrm{Mpc}/h$, the redshift evolution is reversed compared to large scales, i.e., $\omega(r)$ decreases with decreasing redshift along the merger tree~(up to factors of $\sim2$ from $z=3$ to $z=0.6$).   

We compare $\omega(r)$ predictions of \texttt{SAMPLE-TREE} to that of \texttt{SAMPLE-MCUT}; Figure~\ref{fig:comparison} shows the ratio between the two as a function of $r$. We find that as redshift increases, $\omega(r)$ for \texttt{SAMPLE-TREE} becomes increasingly suppressed at scales $r\gtrsim1~\mathrm{Mpc}/h$ as compared to that of \texttt{SAMPLE-MCUT}; at $z=3$ the suppression is by factors $3-4$. At $r\lesssim1~\mathrm{Mpc}/h$, the differences are relatively small~(by factors $\lesssim2$). These differences are largely because \texttt{SAMPLE-TREE} galaxies are less massive compared to \texttt{SAMPLE-MCUT} galaxies at higher redshifts.

In the following subsections, we shall dig deeper into the foregoing results by first putting them in the context of the galaxy-subhalo misalignments, and then finally revealing the factors that drive the evolution of ED correlations at different scales.

%\rachel{Am I right in thinking that the subsubsections below are meant to dig into the results above?  If so, you should clarify this aspect of the narrative.}

\subsubsection{Implications of galaxy-subhalo misalignment on the ED correlation}

We now study the implications of galaxy-subhalo misalignment and its evolution on the ED correlation function. To do this, we compare the ED correlations of galaxies~(also shown in Figure~\ref{fig:corr}) to their underlying dark matter subhaloes. The top panel of Figure~\ref{fig:dm_bm} shows the ED correlation functions of \texttt{SAMPLE-TREE} galaxies, where the solid and dashed lines correspond to galaxies and dark matter subhaloes, respectively. As a consequence of the misalignment between stellar matter and dark matter, the solid lines showing the galaxy ED correlation functions are significantly suppressed compared to the subhalo ED correlation functions~(by factors $\sim2-4$) at all scales. This implies that the alignment of galaxies with respect to the surrounding density field is suppressed as compared to their dark matter subhaloes. This has been established in previous works~\citep{2015MNRAS.453..469T}, and is also supported observationally in the alignments of luminous red galaxies \citep{2009ApJ...694..214O}. %\rachel{There is earlier work on this, e.g., by Teppei Okumura when studying LRG alignments.}
We now discuss how this suppression evolves with redshift on the merger tree. In the bottom panel of Figure~\ref{fig:dm_bm}, we see that the ratio $\omega_{\mathrm{DM subhalo}}/\omega_{\mathrm{galaxy}}$ decreases with decreasing redshift; this is because  galaxy-subhalo misalignment decreases with decreasing redshift. Furthermore, the evolution is stronger for $M_h>10^{11}~M_{\odot}/h$ haloes as compared to $M_h>10^{13}~M_{\odot}/h$. This is because at $z=3$, $M_h>10^{13}~M_{\odot}/h$ galaxies are more aligned with their subhaloes as compared to $M_h>10^{11}~M_{\odot}/h$ galaxies~(compare leftmost and rightmost panels of Figure \ref{fig:theta_hist}). 

\subsubsection{What drives the evolution of ED correlation at different scales?}

Here, we discuss the factors driving the evolution of the galaxy ED correlation at different scales, as inferred from Figure~\ref{fig:dm_bm}. 
At scales $\gtrsim1~\mathrm{Mpc}/h$, note that the ED correlations for dark matter subhaloes~(dashed lines) undergo a significantly weaker redshift evolution compared to that of galaxies~(solid lines). In fact, there is no significant evolution for $M_h>10^{11}~M_{\odot}/h$ and $M_h>10^{12}~M_{\odot}/h$ subhaloes. Therefore, the fact that we find a significant evolution for the galaxy ED correlation implies that its evolution at scales $>1~\mathrm{Mpc}/h$ is primarily driven by the evolution of the galaxy-subhalo misalignment, as opposed to being driven by the ED correlation for dark matter haloes.

At scales $\lesssim1~\mathrm{Mpc}/h$, a suppression~(compared to a power law) is seen in the ED correlations for both galaxies and their dark matter subhaloes. Furthermore, the suppression in the galaxy ED correlation simply traces that of the dark matter subhalo, but at a lower normalization. Overall, this tells us that the evolution of the ED correlation profile for galaxies at scales $\lesssim1~\mathrm{Mpc}/h$ is governed by the evolution of both 1) the ED correlation for dark matter haloes, and 2) the misalignment between galaxies and subhaloes. The former leads to a decrease in the ED correlation for galaxies with time, whereas the latter drives an  increase in the ED correlation for galaxies. Due to the complex interplay between these two competing effects, no straightforward trend is seen in the evolution of ED correlation at scales $\sim 1~\mathrm{Mpc}/h$~(to be targeted by LSST).

At very small scales ($\sim 0.1~\mathrm{Mpc/h}$), the suppressed ED correlation of DM subhaloes is so large that it dominates compared to the evolution of galaxy subhalo misalignment angle. This competition causes the reversal in the redshift evolution of $\omega(r)$ for galaxies at these scales, compared to that in scales $>1~\mathrm{Mpc}/h$.

\section{Conclusions} \label{S:conclusions}

This work is part of a continued series of papers dedicated to studying %the redshift evolution of 
the intrinsic alignments~(IA) of galaxies using the \texttt{MassiveBlackII} cosmological hydrodynamic simulation. In this work, we study redshift evolution~($0.6\lesssim z \lesssim 3$) by selecting galaxy samples (\texttt{SAMPLE-TREE}) based on subhalo mass cuts~($M_h>10^{11,12,13}~M_{\odot}/h$) at $z=0.6$ and tracing their progenitors to $z=3$ along a merger tree. We study the redshift evolution of galaxy shapes, misalignment with respect to host subhalo, and the ED correlation functions along the merger tree. Our key findings are as follows:  
\begin{itemize}
\item The sphericity of the dark matter component of galaxies increases with time, whereas that of the stellar matter component decreases with time.     

\item The distribution of galaxy-subhalo misalignment angle peaks at $\sim$10~deg. With decreasing redshift, the distribution becomes narrower and more skewed towards smaller misalignment angles. 

\item The evolution of the ellipticity-direction~(ED) correlation~$\omega(r)$ of galaxies is driven by the evolution of their alignment with respect to their host DM subhaloes, as well as the alignment between DM subhaloes and the surrounding matter overdensity. 
\begin{itemize}
\item At scales  $\sim1~\mathrm{cMpc}/h$, the alignment between DM subhaloes and the matter overdensity gets suppressed with time. On the other hand, the alignment between galaxies and DM subhaloes is enhanced. Due to these competing tendencies, the redshift evolution of $\omega(r)$ for galaxies at $\sim1~\mathrm{cMpc}/h$ is not straightforward. 

\item At scales $>1~\mathrm{cMpc}/h$, there is no significant evolution in the alignment between DM subhaloes and the matter overdensity. As a result, the evolution of the galaxy-subhalo misalignment leads to an increase in $\omega(r)$ for galaxies by a factor of $\sim$4 from $z=3$ to $0.6$. 

\item At $\sim0.1~\mathrm{cMpc}/h$ scales, evolution in $\omega(r)$ for galaxies is completely reversed compared to that at scales $\gtrsim1~\mathrm{cMpc}/h$, i.e., it decreases by factors $\sim 2$ from $z=3$ to $0.6$. This is because at these scales, the alignment between DM subhaloes and the matter overdensity is strongly suppressed with time, and this effect dominates over evolution of galaxy-subhalo misalignment.  
\end{itemize}
%\begin{itemize}

%\item \aklant{The evolution of galaxy ED correlation amplitude at  scales $\gtrsim 1~\mathrm{cMpc}/h$ is primarily governed by the evolution of the galaxy-subhalo misalignment. This is because the ED correlation of dark matter subhaloes undergoes no significant redshift evolution.}

%\item \aklant{The evolution of galaxy ED correlation amplitude at scales $\lesssim 1~\mathrm{cMpc}/h$ is governed by both, ED correlation of dark matter subhaloes, as well as the galaxy-subhalo misalignment angle. The former drives the evolution of  the shape, and the latter drives the evolution of the relative normalization between the ED correlations of galaxies and their dark matter subhaloes.}   
%\end{itemize}

\end{itemize}

We also compare our results with the sample selection applied in the previous work of this series \citep{tenneti2015intrinsic}. In particular, we also considered galaxy samples~(\texttt{SAMPLE-MCUT}) with fixed subhalo mass cuts~($M_h>10^{11,12,13}~M_{\odot}/h$), applied at all redshifts between 0.6 and 3. 

Interestingly, upon comparing the sphericities of dark matter components of \texttt{SAMPLE-TREE} and \texttt{SAMPLE-MCUT} galaxies, we find that they do not significantly differ~($\lesssim 10\%$); this is true even at the highest redshift~($z=3$) where \texttt{SAMPLE-TREE} galaxies are significantly less massive than \texttt{SAMPLE-MCUT}. This is explained by our finding that at $z\gtrsim1.5$, progenitors of $z\sim0.6$: $M_h\gtrsim10^{11}~M_{\odot}/h$ galaxies have significantly less spherical~(on an average) dark matter shapes than a \textit{randomly selected} galaxy of similar subhalo mass and redshift. 

For the stellar matter component, we find that \texttt{SAMPLE-TREE} progenitors at $z=3$ are less spherical compared to \texttt{SAMPLE-MCUT} galaxies. This is because \texttt{SAMPLE-MCUT} galaxies are more massive (which we show to be more spherical for stellar matter component) than \texttt{SAMPLE-TREE} galaxies at $z=3$.

We find that \texttt{SAMPLE-TREE} galaxies are less aligned with their subhaloes compared to \texttt{SAMPLE-MCUT} galaxies. At $z=1.5$ and $z=3$, this can be attributed to the differences between their subhalo masses. But the fact that we also see this at $z=0.6$ further implies that galaxies which formed earlier than $z=3$~(i.e. those that do not have progenitors up to $z=3$) are more aligned than those that formed at $z<3$.

The effect of differences in subhalo masses~(at $z>0.6$) of \texttt{SAMPLE-TREE} and \texttt{SAMPLE-MCUT} galaxies, is also seen in their ED correlation function $\omega(r)$. Compared to \texttt{SAMPLE-MCUT}, $\omega(r)$ for \texttt{SAMPLE-TREE} galaxies is suppressed at increasing redshift (by factors up to $\sim3-4$ at $z=3$); this is due to decreasing subhalo masses of progenitors in \texttt{SAMPLE-TREE} at increasing redshift.

This work demonstrates that hydrodynamic simulations such as MBII are indispensible tools to study redshift evolution of galaxy properties such as IA, primarily because of the ability to directly trace progenitors of present-day galaxies by constructing merger trees. This enables us to disentangle true IA evolution from apparent evolution due to sample selection effects, which are inevitable in observations.  
Future work will involve the use of the results from this study, as well as previous works \citep{tenneti2014galaxy,tenneti2015intrinsic,tenneti2016intrinsic}, to construct halo models for IA of galaxies. These models can then be used to construct mock catalogs by populating N-body simulation volumes, and thereby analyse possible systematic biases caused by IA in weak lensing analyses.

\section*{Acknowledgements}
We thank Yu Feng for providing the data of MB-II simulation snapshots and raw data.  This research is supported by the US
National Science Foundation under Grant No.\ 1716131.
TDM acknowledge funding from NSF ACI-1614853, NSF AST-1517593, NSF AST-1616168, NASA ATP 80NSSC18K1015
and NASA ATP 17-0123.

% Entry for the table of contents, for this guide only
\addcontentsline{toc}{section}{Acknowledgements}

%%%%%%%%%%%%%%%%%%%%%%%%%%%%%%%%%%%%%%%%%%%%%%%%%%



\bibliographystyle{mnras}
\bibliography{example} % if your bibtex file is called example.bib

\appendix

%Fig A1 shows the normalized histograms of axis ratios at $z = 0.6$ showing a comparison between shapes determined by using all particles in the subhalo with those obtained using a random subsample of 50, 300, 500 and 1000 particles in the subhalo. From this plot, we can confirm that using subhalos containing more than 300 particles reach the convergence, thus we only study subhalos with more than 300 dark matter and stellar matter particles.

\section{Fitting the evolution of shape and misalignment angle} \label{S:appendixb}
\begin{figure*}
 \includegraphics[width=150mm]{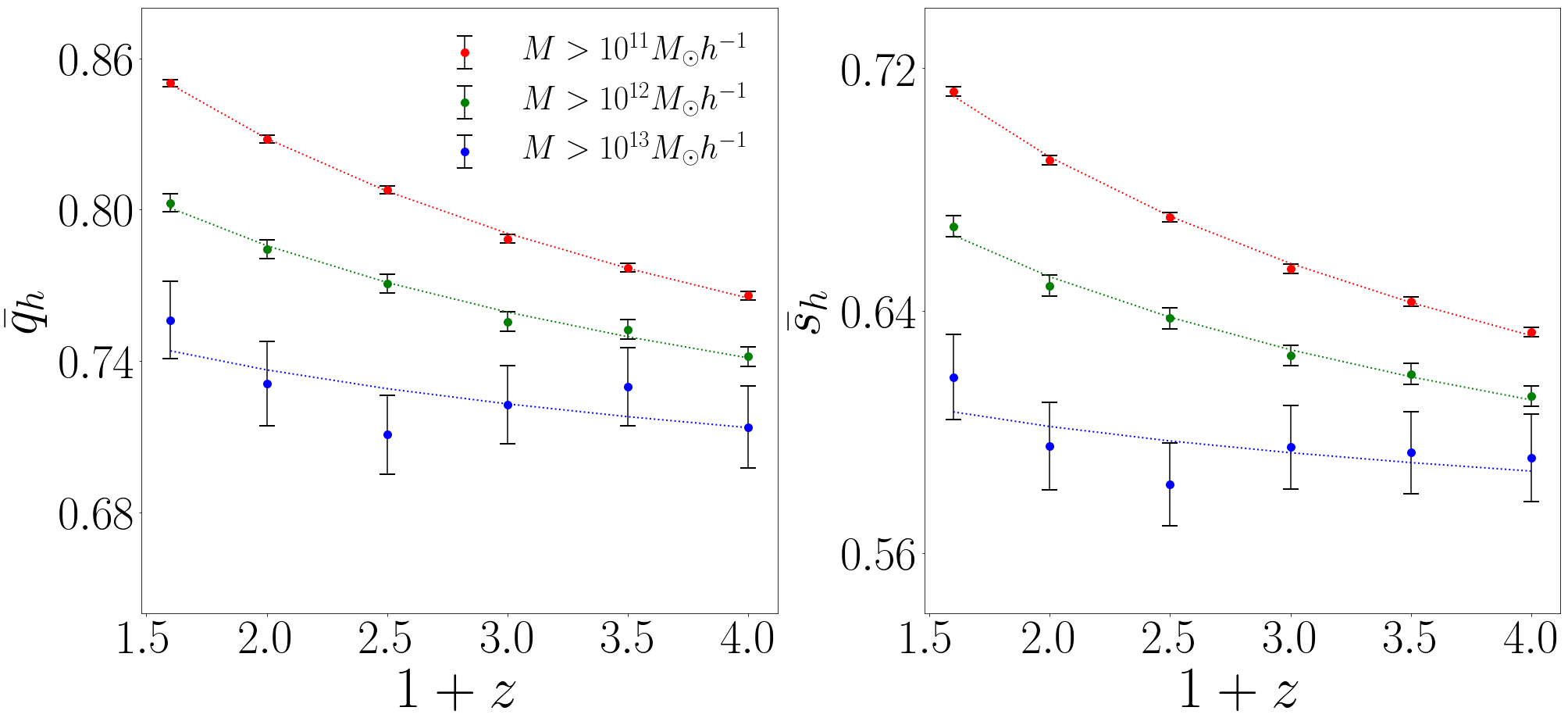}
 \caption{The filled circles show $q_h^{\mathrm{max}}$ and $s_h^{\mathrm{max}}$ which correspond to the averages of the distributions $P(q|M_h,z)$ and $P(s|M_h,z)$ for the dark matter component of \texttt{SAMPLE-TREE} galaxies. The dashed lines show the best fitting trend described by the function in Eq.~(\ref{fitting_equation}. The error bars are jackknife errors obtained by dividing the simulation volume into eight octants.}
 \label{fitting_q_dark}
\end{figure*}

\begin{figure*}
 \includegraphics[width=150mm]{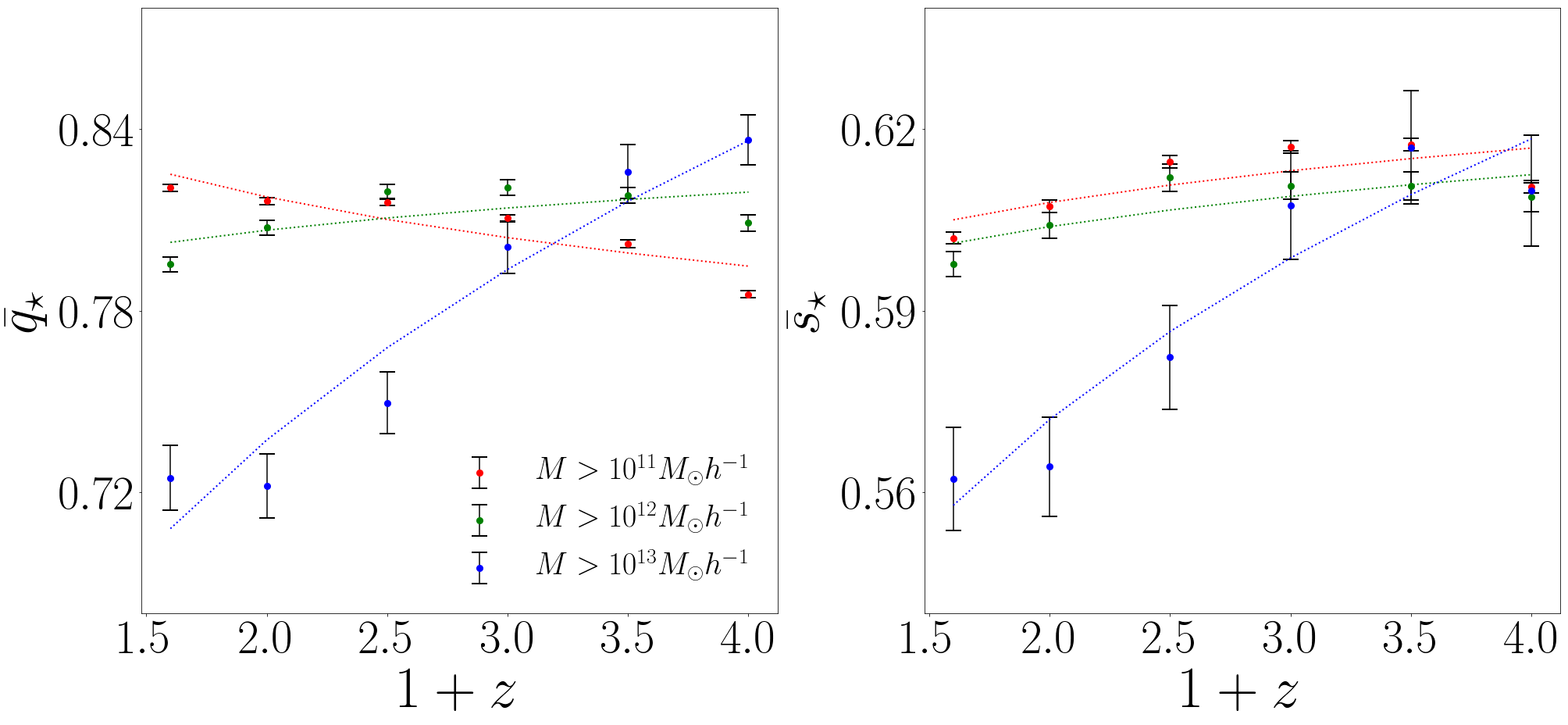}
 \caption{The filled circles show $q_*^{\mathrm{max}}$ and $s_*^{\mathrm{max}}$ which correspond to the averages of the distributions $P(q|M_h,z)$ and $P(s|M_h,z)$ for the stellar matter component of \texttt{SAMPLE-TREE} galaxies. The dashed lines show the best fitting trend described by the function in Eq.~\eqref{fitting_equation}. The error bars are jackknife errors obtained by dividing the simulation volume into eight octants.}
 \label{fitting_q_stellar}
\end{figure*}

\begin{figure}
 \includegraphics[width=80mm]{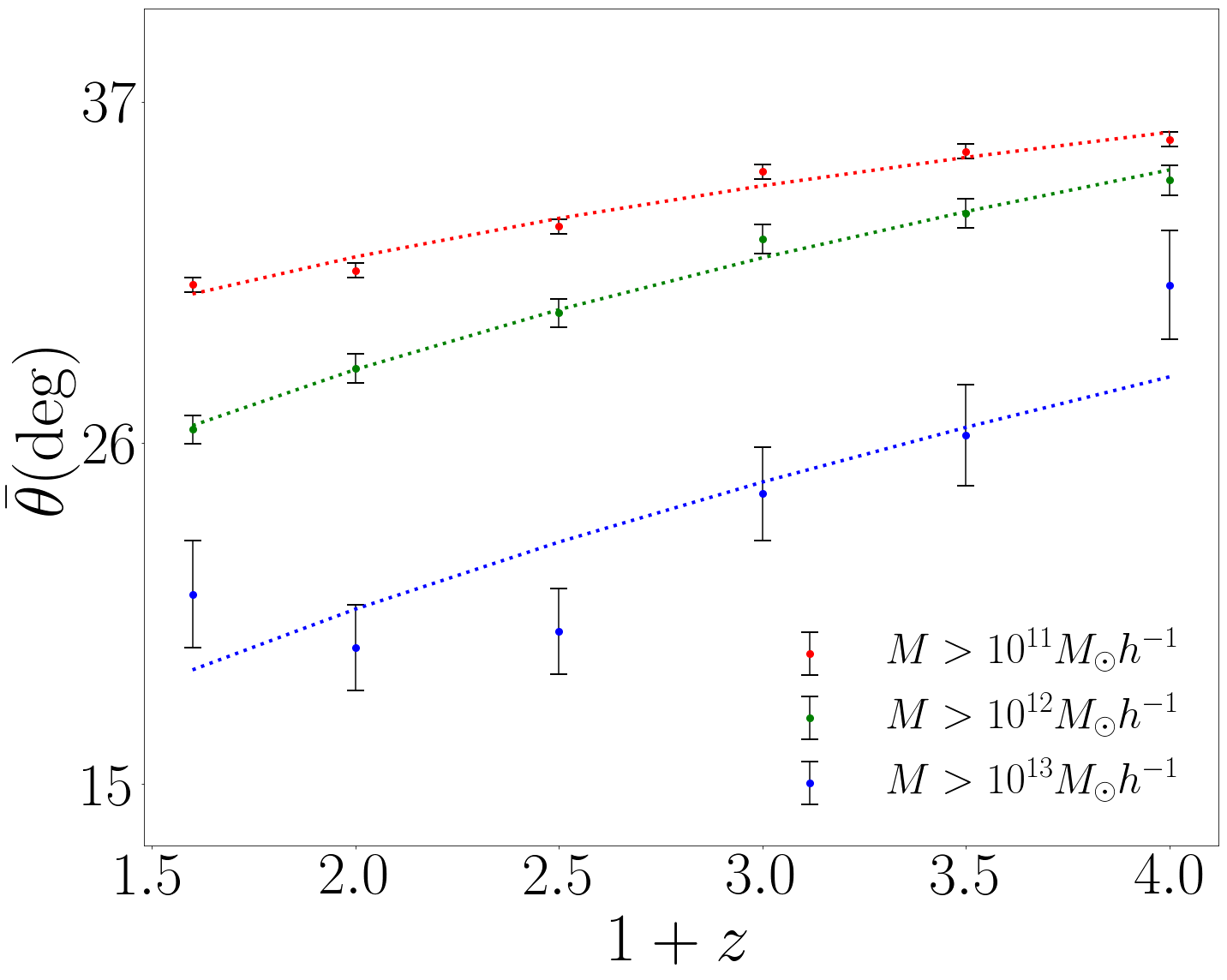}
 \caption{The filled circles show $\theta^{\mathrm{max}}$ which corresponds to the peaks of the distributions $P(\theta|M_h,z)$ for the dark matter component of \texttt{SAMPLE-TREE} galaxies. The dashed lines show the best fitting trend described by the function in Eq.~\eqref{fitting_equation}.}
 \label{fitting_angle}
\end{figure}

Here we present fitting functions for the evolution of the axis ratios and misalignment angles. We model the redshift evolution as a power-law, 
\begin{center}
\begin{eqnarray}
\bar{X}=X_0 (1+z)^{-\alpha_X}
%q=s_0(1+z)^{-\alpha}
\label{fitting_equation}
\end{eqnarray}
\end{center}
where $X$ represent the quantity of interest (axis ratios/ misalignment angle). $\bar{X}$ is the average value of the distribution $P(X|M_H,z)$. $X_0$ is the value of $\bar{X}$ at $z=0$. The filled circles in figures \ref{fitting_q_dark}, \ref{fitting_q_stellar} and \ref{fitting_angle} correspond to the average of the distributions of axis ratios and misalignment angles. The dashed lines show the best fits obtained using Eq.~(\ref{fitting_equation}). The corresponding best fit parameters are shown in Table~\ref{table_fits}1.

\begin{table}
%\caption{Parameters for fitting results of \texttt{SAMPLE-TREE} under $M_h>10^{11}M_{\odot}h^{-1}$}
\centering
\begin{tabular}{|l|c|r|}
 \hline
X & $X_0$ & $\alpha_X$\\
 \hline
$q_{h}$ & 0.897 & 0.115 \\
 \hline
$s_{h}$ & 0.756 & 0.129 \\
 \hline
$q_{\star}$ & 0.841 & 0.04 \\
 \hline
$s_{\star}$ & 0.599 & -0.02 \\
 \hline
$\theta$ & 28.4 & -0.170 \\
 \hline
\label{table_fits}
\end{tabular}

\caption{Best fit values of $X_0$ and $\alpha_X$ for various quantities: $q_h$ and $s_h$ are $q$ and $s$ values of dark matter and stellar matter component of the \texttt{SAMPLE-TREE} galaxies~(subhaloes) respectively.}
\end{table}

%\section{Dark matter particle tracers for calculating ED correlation function}

%In order to take the effect of subhalo bias at large scales into consideration, we considered the ratio of two-point correlation functions using the dark matter particles to trace the density field with those obtained by using the subhalos to trace the density field.

%Fig C1 shows the ratio plot of correlation function of galaxies on the merger tree using dark matter density tracers versus galaxy number density tracers at different redshifts $z=0.6$, $z=1.5$ and $z=3.0$ of \texttt{SAMPLE-TREE}. We can see that the correlation using dark matter density tracers are greater than using galaxy number density tracers. At small scales, the ratio increases with decreasing redshift.

%\begin{figure*}
% \includegraphics[width=\textwidth]{par_gal_ratio.png}
%% \label{fig:par_gal_ratio}
%\end{figure*}

%%%%%%%%%%%%%%%%%%%%%%%%%%%%%%%%%%%%%%%%%%%%%%%%%%

% Don't change these lines
\label{lastpage}

\end{document}